\documentclass[english,12pt,aps,prd,a4paper,preprintnumbers,floatfix,nofootinbib,showpacs,superscriptaddress, notitlepage]{revtex4-1} 
\usepackage[mode=buildnew]{standalone}
\usepackage[usenames,dvipsnames]{color}  
\usepackage{graphicx}
\usepackage{setspace}
\usepackage{caption}
\usepackage{amsmath}
\usepackage{amssymb}
\usepackage[colorlinks=true,citecolor=darkred,urlcolor=darkred, pdfborder={0 0 0}]{hyperref}
\usepackage[normalem]{ulem}
\usepackage{xcolor}
\usepackage{placeins}
\usepackage{mathrsfs}
\usepackage{verbatim} 
\usepackage{cancel} 
\usepackage{float} 
\usepackage{multirow}
\usepackage{scalerel}
\usepackage{tikz-feynman}
\tikzfeynmanset{compat=1.1.0} 
\usepackage{feynmp}
\usepackage{tikzsymbols}
\usepackage{subfigure}

\makeatletter
\def\p@subsection{}
\makeatother
\definecolor{darkred}{rgb}{0.6,0,0}

\definecolor{linkcolor}{rgb}{0,0,0.5}

\usepackage[T1]{fontenc} 
\usepackage{float}

\usepackage[compat=1.1.0]{tikz-feynhand}
\usepackage{tikz-feynman}
\tikzfeynmanset{compat=1.1.0}
\usepackage{feynmp}
\usepackage{tikzsymbols}
\usepackage{array}
\usepackage{pifont} 
\usepackage{amsmath}
\usepackage{mathrsfs}
\usepackage{placeins}

\definecolor{linkcolor}{rgb}{0,0,0.5}

\def\0nbb {$0\nu\beta\beta$ }

\def\Z2{$\mathcal{Z}_2$}
\def\A4{$A_4$}

\usepackage{colortbl}
\usepackage{nicefrac}
\usepackage{nicematrix}

\usepackage{booktabs}

\newcommand{\AddrSeoultech}{Seoul National University of Science and Technology, Seoul 01811, Republic of Korea}

\begin{document}

\title{\color{BrickRed} Radiative Dirac Neutrino Masses from Modular $S_3$ Symmetry in an Axion Model}

\author{Sin Kyu Kang}
\email{skkang@seoultech.ac.kr}
\affiliation{\AddrSeoultech}
\author{Ranjeet Kumar}
\email{kumarranjeet.drk@gmail.com}
\affiliation{Institute for Convergence of Basic Studies, Seoul National University of Science and Technology, Seoul 01811, Republic of Korea}
\author{Hiroshi Okada}
\email{hiroshi3okada@htu.edu.cn}
\affiliation{Department of Physics, Henan Normal University, Xinxiang 453007, China}

\begin{abstract}
\vspace{1cm}
\noindent
We present a unified axion model framework that simultaneously addresses the origin of neutrino masses, the leptonic flavor structure, the strong CP problem, and dark matter. The model is based on a global $U(1)_{\rm PQ}$ symmetry combined with a modular $S_3$ symmetry and is realized within a novel class of KSVZ-type axion model. Exotic colored fermions and scalars mediate radiative neutrino mass generation at the one loop-level. The PQ charge assignment forbids tree-level neutrino masses and leaves a residual $Z_3$ symmetry that ensures the Dirac nature of neutrinos. In the minimal realization, the neutrino mass matrix is of rank two, predicting one massless neutrino. Consequently, the sum of neutrino masses is constrained for both the normal and inverted hierarchies. We analyze the implications for charged lepton flavor violation and the lepton $g-2$. The axion emerging from this framework dynamically resolves the strong CP problem and accounts for the observed dark matter abundance. Notably, the predicted axion-photon coupling is within reach of upcoming experiments and consistent with existing astrophysical and cosmological bounds. 
\end{abstract}

\maketitle

\section{Introduction} \label{sec:intro}
The experimental confirmation of neutrino oscillations has firmly established that neutrinos possess nonzero masses and undergo flavor mixing \cite{Kamiokande-II:1990wrs,Kamiokande-II:1992hns,Super-Kamiokande:1998kpq,Cleveland:1998nv,SNO:2002tuh}. At the same time, a wide range of astrophysical and cosmological observations indicate that nearly $85\%$ of the matter content of the Universe is composed of dark matter (DM) \cite{Planck:2018vyg}. These observations highlight the limitations of the Standard Model (SM), which lacks both a mechanism for neutrino mass generation and a viable DM candidate and thus points to physics beyond the SM (BSM). Therefore, extensions of the SM that simultaneously address neutrino masses and DM within a single framework are particularly appealing. Among such approaches, the scotogenic model~\cite{Ma:2006km} provides a simple and elegant framework in which neutrino masses arise radiatively and are intimately connected to DM phenomenology. Numerous variants of scotogenic models have been proposed, realizing neutrinos as either Majorana or Dirac particles~\cite{Kumar:2019tat,Guo:2020qin,Leite:2020wjl,Kang:2021jmi,Nomura:2021aep,Borah:2022enh,Borah:2022phw,ChuliaCentelles:2022ogm,Nomura:2023vmh,Singh:2023eye,Kumar:2024zfb,Bharadwaj:2024crt,Borah:2024gql,CentellesChulia:2024iom,Nomura:2024vzw,Nomura:2024zca,Singh:2025jtn,Lozano:2025tst,Hundi:2025reg,Guo:2025xmz,Kumar:2025aek,Avila:2025qsc,Kumar:2025cte,AbuSiam:2025voc,Nomura:2025ovm,Nomura:2025yoa,Jobu:2025tto,Kumar:2025zvv,Nasri:2026nbf,Kang:2026lgr}.

Another longstanding puzzle in particle physics is the apparent absence of CP-violation in Quantum Chromodynamics (QCD). Experimental constraints require the CP-violating parameter $\bar{\theta}$ to be extremely small, $|\bar{\theta}| < 10^{-10}$ \cite{Baker:2006ts,Pendlebury:2015lrz}. This tension constitutes the strong CP problem, raising the fundamental question of why CP symmetry is strongly violated in the weak interactions while it appears to be preserved in the strong force. The strong CP problem can be naturally resolved within the Peccei-Quinn (PQ) framework \cite{Peccei:1977hh,Peccei:1977ur} through the introduction of a new global symmetry denoted as $U(1)_{\rm PQ}$. The spontaneous breaking of this symmetry gives rise to a pseudo-Goldstone boson known as the axion \cite{Weinberg:1977ma,Wilczek:1977pj}. Once the PQ symmetry is spontaneously broken, it induces an effective term with the same structure as the QCD $\bar{\theta}$-term. As a result, the CP-violating phase $\bar{\theta}$ is promoted to a dynamical axion field, whose vacuum minimum effectively leads $\bar{\theta}$ to be zero, thereby resolving the strong CP problem. Well studied axion models such as KSVZ \cite{Kim:1979if,Shifman:1979if} and DFSZ \cite{Zhitnitsky:1980tq,Dine:1981rt} provide distinct realizations of the PQ mechanism, characterized by different PQ charge assignments to SM quarks and additional heavy fermions. 

Most theoretical efforts have focused on resolving these puzzles separately by introducing distinct new particles or symmetries for each phenomenon. However, recently, a KSVZ-type axion model has been proposed that simultaneously addresses neutrino mass generation, the strong CP problem, and the DM phenomenology within a unified framework~\cite{Batra:2025gzy}. In this setup, neutrino masses are generated radiatively at the one loop-level, following a mechanism analogous to that of the scotogenic model~\cite{Ma:2006km}. A similar idea has also been explored for the two-loop neutrino mass generation~\cite{Batra:2023erw,Kang:2024jnp}. 
These constructions also exploit the fact that axion can also serve as a viable cold DM candidate~\cite{Preskill:1982cy,Abbott:1982af,Dine:1982ah}. 

In this work, we introduce a KSVZ-type axion model framework in which neutrino mass generation, the strong CP problem, and DM phenomenology are addressed in a unified manner at the one loop-level. Additionally, the leptonic flavor structure emerges from an underlying modular $S_3$ symmetry, leading to a symmetry motivated explanation of neutrino mixing.\footnote{Modular symmetry has recently emerged as a compelling framework, in which the usual discrete flavor symmetries are generalized to modular symmetries~\cite{Feruglio:2017spp}, where Yukawa couplings are promoted to modular forms that transform non-trivially under the modular group, resulting in predictive lepton and/or quark mixing patterns~\cite{Kobayashi:2018vbk,Meloni:2023aru,Okada:2019xqk,Mishra:2020gxg,Belfkir:2024wdn,Behera:2024ark,Behera:2025tpj,Tavartkiladze:2025oiq,Nomura:2023usj,Mishra:2022egy,Wang:2019xbo,Kashav:2021zir,Kashav:2022kpk,Mishra:2023ekx,Lu:2019vgm,RickyDevi:2024ijc, Kobayashi:2019gtp,Nomura:2019xsb,Behera:2020lpd,MiteshKumar:2023hpg,Nomura:2023kwz,Kim:2023jto,Du:2022lij,Devi:2023vpe,Kashav:2021zir,Dasgupta:2021ggp,Nomura:2019lnr,Kashav:2022kpk,CentellesChulia:2023zhu,Kumar:2023moh,Devi:2023vpe,Kobayashi:2023qzt,Nomura:2024ghc,Nomura:2024ctl,Nomura:2024vus,Singh:2024imk,Pathak:2024sei,Nomura:2025bph,CentellesChulia:2025bcg,Kumar:2026qee,Pathak:2025zdp,Penedo:2018nmg,Novichkov:2018ovf,Kobayashi:2019xvz,Liu:2020akv,deMedeirosVarzielas:2023crv,Ding:2021zbg,King:2019vhv,Novichkov:2018nkm,Ding:2019zxk}.} We have introduced the vector-like fermions ($\Psi_{L_i}, \Psi_{R_i}$), scalar $\chi$, singlets under $SU(2)_L$ and scalar $\eta$, doublet of $SU(2)_L$. 
They are triplets under the fundamental representation of $SU(3)_C$
and singlets of modular $S_3$ symmetry carrying $-2$ modular weight. In addition, we consider a global $U(1)_{\rm PQ}$ symmetry, whose spontaneous breaking leads to the axion. This provides a solution to the strong CP problem as well as a promising axion DM candidate. The $U(1)_{\rm PQ}$ symmetry is spontaneously broken by the vacuum expectation value (vev) of the newly added complex scalar $\sigma$. Except for the Higgs field $H$, all fields in the model carry non-trivial PQ charges. Moreover, the PQ charge assignment forbids neutrino masses at the tree level, and the breaking of $U(1)_{\rm PQ}$ leads to a residual $Z_3$ symmetry, which ensures the Dirac nature of neutrinos. We consider the effective operator $\overline{L}\tilde{H}\nu_{R}\sigma^{\ast}$ for Dirac neutrino mass generation, whose ultraviolet (UV) completion is realized by the colored fields, $\Psi_{L_i}, \Psi_{R_i}$, $\chi$, and $\eta$ at the one loop-level. 
The PQ breaking scale, determined by the vev of $\sigma$, governs the masses of $\Psi_i$, $\chi$, and $\eta$. Since these fields mediate the neutrino mass generation at the one loop-level, the origin of neutrino masses becomes intimately tied to the PQ breaking scale and consequently to the axion mass.

The remainder of the paper is organized as follows. In Sec.~\ref{sec:model}, we introduce our model framework and discuss the scalar sector and its mass spectrum, as well as derive the charged lepton and neutrino mass matrices. We discuss the numerical analysis in Sec.~\ref{sec:numericalanalysis} and present our model's predictions for both the normal hierarchy (NH) and inverted hierarchy (IH) of neutrino masses. A detailed discussion of the axion as a solution to the strong CP problem, a viable DM candidate, and axion-photon coupling is provided in Sec.~\ref{sec:axion}. In Sec.~\ref{sec:lfv}, we analyze the radiative and axion-emitting charged lepton flavor violation (cLFV) and the lepton $g-2$ based on our numerical analysis. Finally, we give our concluding remarks in Sec.~\ref{sec:conc}.

\section{Model Framework} \label{sec:model}
We consider a model framework where Dirac neutrino mass generation is interconnected with the axion. The effective operator $\overline{L}\tilde{H}\nu_{R}\sigma^{\ast}$ has been considered in conjunction with modular $S_3$ symmetry for the Dirac neutrino mass generation and mixing. Similar to the original KSVZ models \cite{Kim:1979if,Shifman:1979if}, we introduce vector-like fermions $\Psi_{L}$ and $\Psi_{R}$. These are triplets in the fundamental representation of $SU(3)_C$, singlets under $SU(2)_L$, and have hypercharge $Y = 0$. In the minimal realization of the model, we consider two generations of $\Psi_{L_i}$ and $ \Psi_{R_i}$ ($i=1,2$). As a consequence, only two neutrinos acquire nonzero masses, while one remains massless, consistent with neutrino oscillation data \cite{Kamiokande-II:1990wrs,Kamiokande-II:1992hns,Super-Kamiokande:1998kpq,Cleveland:1998nv,SNO:2002tuh}. 
We further incorporate a $U(1)_{\rm PQ}$ symmetry, which serves multiple purposes within our model. The $U(1)_{\rm PQ}$ charge assignment of the left handed lepton doublet $L$ and the right handed neutrino $\nu_R$ forbids the tree-level mass terms. In addition, the $U(1)_{\rm PQ}$ charges have been chosen in such a way that the breaking of $U(1)_{\rm PQ}$ symmetry leads to the residual $Z_3$ symmetry: $U(1)_{\rm PQ} \to Z_3$. This residual symmetry ensures the Dirac nature of neutrinos and forbids Majorana mass terms. 
This can be understood from the specific charge assignments of the fields. After rescaling the $U(1)_{\rm PQ}$ charges by the largest denominator factor, namely $6$, the fields $\sigma$, $H$, $L_i$, and $\nu_{R_i}$ acquire charges $3$, $0$, $1$, and $4$, respectively. Consequently, under the residual $Z_3$ symmetry, their transformations are given by $\omega^3$, $\omega^0$,  $\omega$, $\omega^4 \equiv 1$, $1$, $\omega$, $\omega$, respectively. As a result, the Majorana type terms $\bar{L}_i^c L_j$ and $\bar{\nu}_{R_i}^c \nu_{R_j}$ are not invariant under $Z_3$ and therefore remain forbidden, whereas the Dirac type operator $\overline{L}\tilde{H}\nu_R \sigma^\ast$ is $Z_3$ invariant and hence allowed.
Earlier, this has been explored in the context of $U(1)_{B-L} \to Z_N$ \cite{Bonilla:2018ynb,CentellesChulia:2019gic,Srivastava:2019xhh,Borboruah:2024lli,Kumar:2025cte}. The spontaneous breaking of $U(1)_{\rm PQ}$ is achieved via a newly added complex scalar singlet $\sigma$, which also generates masses for the vector-like fermions. The phase of the scalar $\sigma$ corresponds to the axion field $a$. Assigning different $\rm{PQ}$ charges to $\Psi_{L_i}$ and $\Psi_{R_i}$ ensures the anomalous axion-gluon coupling responsible for addressing the strong $\rm CP$ problem.

These vector-like fermions also serve as mediators for neutrino mass generation at the one loop-level. To UV-complete the operator $\overline{L} \tilde{H} \nu_{R}\sigma^{\ast}$, two additional scalars, $\eta$ and $\chi$, are introduced, which are triplets under the fundamental representation of $SU(3)_C$. The scalar $\eta$ ($\chi$) is an $SU(2)_L$ doublet (singlet) with $Y=1/2 $ ($Y=0$). The $\rm PQ$ charge assignments of the fields propagating in the loop are parametrized by a free parameter $q$, which denotes the charge of $\Psi_{R_i}$. Furthermore, under the modular $S_3$ symmetry, only leptons $L_{23} \equiv (L_2, L_3)^T$, $e_{R_{23}} \equiv (e_{R_2}, e_{R_3})^T$, and right handed neutrino $\nu_{R_{23}} \equiv (\nu_{R_2}, \nu_{R_3})^T$ are assigned doublets ($\mathbf{2}$), while $\Psi_{L_2}$, $\Psi_{R_2}$ are singlets ($\mathbf{1'}$), and all the remaining fields of the model are assigned trivial singlet ($\mathbf{1}$) under $S_3$. The modular weight assignments are such that $e_{R_{23}}$, $\nu_{R_{23}}$, $H$, and $\sigma$ are assigned weight $0$, while the remaining fields have weight $-2$. The field contents and their transformation properties under the different symmetries, along with the modular weights, are summarized in Table~\ref{tab:fields}.
\begin{table}[!h]
\centering
\small
\setlength{\tabcolsep}{4pt}
\renewcommand{\arraystretch}{1.2}
\begin{tabular}{|c|cccccccccc|cccc|}
\hline
 & \multicolumn{10}{c|}{\textbf{Fermions}} & \multicolumn{4}{c|}{\textbf{Bosons}} \\
 \hline
 & $L_{1}$ & $L_{23}$ & $e_{R_1}$ & $e_{R_{23}}$ & $\nu_{R_1}$& $\nu_{R_{23}}$& $\Psi_{L_1}$ & $\Psi_{L_2}$ & $\Psi_{R_1}$ & $\Psi_{R_2}$ 
 & $H$ & $\eta$ & $\sigma$ & $\chi$  \\
\hline
$SU(3)_C$ 
& 1 & 1 & 1 & 1 & 1 & 1& 3 & 3 & 3 & 3 
& 1 & 3 & 1 & 3  \\

$SU(2)_L$ 
& 2 & 2 & 1 & 1 & 1 &1 & 1 & 1 & 1 & 1 
& 2 & 2 & 1 & 1  \\

$U(1)_Y$ 
& $-\frac{1}{2}$ & $-\frac{1}{2}$ & $-1$ & $-1$ &0 &0 & 0 & 0 & 0 & 0 
& $\frac{1}{2}$ & $\frac{1}{2}$ & 0 & 0  \\

$U(1)_{\rm PQ}$ 
& $\frac{1}{6}$ & $\frac{1}{6}$  & $\frac{1}{6}$  & $\frac{1}{6}$ & $\frac{2}{3}$ &$\frac{2}{3}$& $q+\frac{1}{2}$  & $q+\frac{1}{2}$ & $q$ & $q$ 
& 0 & $q-\frac{1}{6}$ & $\frac{1}{2}$ & $q-\frac{1}{6}$  \\

$S_3$ 
& 1 & 2 & 1 & 2 & 1&2 & 1 & $1'$ & 1 & $1'$ 
& 1 & 1 & 1 & $1$ \\

$-k$ 
&$\overline{-2}$ & $\overline{-2}$ & $-2$ & 0 &$-2$ &$0$ & $\overline{-2}$ & $\overline{-2}$ & $-2$ & $-2$ 
& 0 & $\overline{-2}$ & 0 & $-2$  \\
\hline
\end{tabular}
\caption{Field content and transformation properties of different fields under the
$SU(3)_C \otimes SU(2)_L \otimes U(1)_Y \otimes U(1)_{\rm PQ} \otimes S_3 $ symmetry.
Here, $-k$ denotes the modular weight and $\overline{-2}$ for the left handed fermions $\psi's\equiv\{L_1,L_{23},\Psi_{L_1},\Psi_{L_2}  \}$ and $\eta$ is defined so that $-2$ is for $\overline {\psi's}$ and $\tilde{\eta}$.
} 
\label{tab:fields}
\end{table}

In what follows, we construct the most general renormalizable Lagrangian consistent with $SU(3)_C\otimes SU(2)_L\otimes U(1)_Y$, the global $U(1)_{\rm PQ}$, and the modular $S_3$ symmetry. 
The $\tau$-dependence of selected parameters is treated effectively, motivated by modular invariance, without committing to a specific UV-complete supersymmetric model.

\subsection{Scalar Sector and the Resulting Mass Spectrum}
Following the charge assignments of the fields under the various symmetries given in Table \ref{tab:fields}, the scalar potential of the model can be formulated as
\begin{align}
\mathcal{V}= &- \mu^2_H |H|^2+ \boldsymbol{\mu^2_{\eta}} |\eta|^2 + \boldsymbol{\mu^2_{\chi}} |\chi|^2 - \mu^2_{\sigma} |\sigma|^2 +  \lambda_{H} |H|^4 +  \lambda_{\sigma} |\sigma|^4 +  \lambda_{H \sigma}  |H|^2 |\sigma|^2
\nonumber \\ &
+   
\boldsymbol{\lambda_{\eta}} |\eta|^4 +  \boldsymbol{\lambda_{\chi}} |\chi|^4 + \boldsymbol{\lambda_{H \eta}} |H|^2 |\eta|^2 +  \boldsymbol{\lambda'_{H \eta}} |H^\dagger \eta|^2 + \boldsymbol{\lambda_{H \chi}}  |H|^2 |\chi|^2 
\nonumber \\ &+
  \boldsymbol{\lambda_{\eta \sigma} }  |\eta|^2 |\sigma|^2 +  \boldsymbol{\lambda_{\chi \sigma}} |\chi|^2 |\sigma|^2  + \boldsymbol{\lambda_{\eta \chi} } 
|\eta|^2 |\chi|^2 
+ \left(
 \kappa Y^{(4)}_{\mathbf{1}}  
\eta ^\dagger H \chi + \text{h.c.} \right),
\end{align}
where all the couplings written in \textbf{bold} have the following definition $\boldsymbol{\Lambda} \equiv \tilde{\Lambda}/(-i \tau + i \bar{\tau})^2$ and their invariance under modular symmetry is similar to the invariance of the K$\ddot{a}$hler potential~\cite{Feruglio:2017spp}. 
The vev of the complex scalar, $\langle \sigma \rangle = v_{\sigma}/\sqrt{2}$, is responsible for the spontaneous breaking of the $U(1)_{\rm PQ}$ symmetry, while the Higgs vev, $\langle H \rangle = \left( 0, \ v_{H}/\sqrt{2} \right)^T$, breaks the electroweak symmetry. After symmetry breaking, the mass spectrum of the scalar sector is obtained. The color neutral scalars $H$ and $\sigma$ mix with each other, whereas the colored scalars $\eta$ and $\chi$ mix among themselves. The corresponding mass matrices are given below.
\begin{align}
    &\mathcal{M}^2_{H \sigma} = \begin{pmatrix}
        2 \lambda_H v_H^2 & \lambda_{H \sigma} v_H v_{\sigma} \\
       \lambda_{H \sigma} v_H v_{\sigma}   & 2 \lambda_{\sigma} v_{\sigma}^2 
    \end{pmatrix},  
    \nonumber \\
    &\mathcal{M}^2_{\eta \chi} = \begin{pmatrix}
        m^2_{\eta}  &   \kappa Y^{(4)}_{\mathbf{1}}  v_H /\sqrt{2}\\
     \kappa Y^{(4)}_{\mathbf{1}}  v_H /\sqrt{2} & m^2_{\chi}
    \end{pmatrix} ,
\end{align}
\begin{align*}
\text{where} \quad \quad &m^2_{\eta} \equiv  \boldsymbol{\mu^2_{\eta}} + \frac{1}{2}\left(\boldsymbol{\lambda_{H \eta}} +  \boldsymbol{\lambda'_{H \eta}}  \right) v_H^2 + \frac{1}{2}\boldsymbol{\lambda_{\eta \sigma}}  v_{\sigma}^2 \ , \\
&m^2_{\chi} \equiv  \boldsymbol{\mu^2_{\chi}} +  \frac{1}{2}\boldsymbol{\lambda_{H \chi}}  v_H^2 + \frac{1}{2} \boldsymbol{\lambda_{\chi \sigma}}  v_{\sigma}^2 \ .
\end{align*}
After electroweak symmetry breaking, only the electrically neutral component of $\eta$ mixes with $\chi$, while $\eta^{\pm}$ remains unmixed. 
Without loss of generality, we absorb the overall phase of $\kappa Y^{(4)}_1$ by field redefinitions and take the $\eta$-$\chi$ mixing parameter to be real.
The scalar fields $\eta$ and $\chi$ are rotated into the mass eigenstate basis $\zeta_1$ and $\zeta_2$ by a rotation matrix $U$ that diagonalizes the corresponding mass matrix, given by
\begin{align} \label{eq:scalarrot}
    \begin{pmatrix}
        \eta \\
        \chi
    \end{pmatrix} = U. \begin{pmatrix}
        \zeta_1 \\
        \zeta_2
    \end{pmatrix}= \begin{pmatrix}
        \cos{\theta} & \sin{\theta} \\
        -\sin{\theta} & \cos{\theta} 
         \end{pmatrix}
         \begin{pmatrix}
            \zeta_1 \\
        \zeta_2
    \end{pmatrix}, \quad  \tan{2 \theta} = \frac{\sqrt{2} \kappa Y^{(4)}_{\mathbf{1}}   v_H}{m^2_{\eta}-m^2_{\chi}} \ .
\end{align}
The resulting physical masses are given by,
\begin{align} \label{eq:scalarmass}
    m^2_{1, 2}=\frac{1}{2} \left(  m^2_{\eta}+m^2_{\chi} \pm \sqrt{\left(m^2_{\eta}-m^2_{\chi}\right)^2 + 2 \left( \kappa Y^{(4)}_{\mathbf{1}}  \right)^2  v^2_H } \right).
\end{align}
Once the $U(1)_{\rm PQ}$ and electroweak symmetries are broken, the remaining fields also acquire masses, in particular the neutrinos, as discussed next.

\vspace{-0.5cm}
\subsection{Leptonic Yukawa Sector and Neutrino Mass Generation} \label{sec:lepton}
 Using the charge assignments under the different symmetries shown in Table~\ref{tab:fields}, the renormalizable Yukawa Lagrangian that governs the leptonic sector can be constructed as follows
\begin{align}
    -\mathcal{L}_y&= \alpha_{\ell}\left( Y^{(4)}_{\mathbf{1}} \otimes \overline{L}_1 \otimes e_{R_1} \right)_1 H + \beta_{\ell}\left( Y^{(2)}_{\mathbf{2}} \otimes \overline{L}_1 \otimes e_{R_{23}} \right)_1 H + \gamma_{\ell}\left( Y^{(4)}_{\mathbf{2}} \otimes \overline{L}_{23} \otimes e_{R_1} \right)_1 H \nonumber \\ &+ \delta_{\ell}\left( Y^{(2)}_{\mathbf{2}} \otimes \overline{L}_{23} \otimes e_{R_{23}} \right)_1 H 
    + \alpha_{\nu}\left( Y^{(6)}_{\mathbf{1}} \otimes \overline{L}_1 \otimes \Psi_{R_1} \right)_1 \tilde{\eta} + \beta_{\nu}\left( Y^{(6)}_{\mathbf{1'}} \otimes \overline{L}_1 \otimes \Psi_{R_{2}} \right)_1 \tilde{\eta} \nonumber \\ &+\gamma_{\nu}\left( Y^{(6)}_{\mathbf{2}} \otimes \overline{L}_{23} \otimes \Psi_{R_1} \right)_1 \tilde{\eta} + \delta_{\nu}\left( Y^{(6)}_{\mathbf{2}} \otimes \overline{L}_{23} \otimes \Psi_{R_{2}} \right)_1 \tilde{\eta} + \alpha'_{\nu}\left( Y^{(6)}_{\mathbf{1}} \otimes \overline{\Psi}_{L_1} \otimes \nu_{R_1} \right)_1 \chi \nonumber \\ & + \beta'_{\nu}\left( Y^{(4)}_{\mathbf{2}} \otimes \overline{\Psi}_{L_1} \otimes \nu_{R_{23}} \right)_1 \chi +\gamma'_{\nu}\left( Y^{(6)}_{\mathbf{1'}} \otimes \overline{\Psi}_{L_2} \otimes \nu_{R_1} \right)_1 \chi + \delta'_{\nu}\left( Y^{(4)}_{\mathbf{2}} \otimes \overline{\Psi}_{L_2} \otimes \nu_{R_{23}} \right)_1 \chi \nonumber \\ &+
    \alpha_{\psi} \left( Y^{(4)}_{\mathbf{1}} \otimes \overline{\Psi}_{L_1} \otimes \Psi_{R_1} \right)_1 \sigma 
+     \delta_{\psi} \left( Y^{(4)}_{\mathbf{1}} \otimes \overline{\Psi}_{L_2} \otimes \Psi_{R_2} \right)_1 \sigma + \rm{h.c.},
\label{eq:yuklag}
\end{align}
where $\tilde{\eta}\equiv i \sigma_2 \eta^\ast$; $\sigma_2$ is the second Pauli matrix.
The lower indices in parentheses, $1$ in the above equation, denote trivial singlet representations under the $S_3$ symmetry.
We provide the relevant modular Yukawa couplings of the model and their weights in the following Table \ref{tab:modyuk} (see App.~\ref{sec:s3rule} for details).
\begin{table}[!h]
    \centering
    \begin{tabular}{|c||c|c|c|c|c|c|c|c|}
    \hline
 & \multicolumn{6}{c|}{\textbf{Couplings}}  \\
 \hline 
         & ~$Y^{(2)}_{\mathbf{2}}$ ~& ~$Y^{(4)}_{\mathbf{1}}$ ~&~ $Y^{(4)}_{\mathbf{2}}$ ~&~ $Y^{(6)}_{\mathbf{1}}$ ~&~$Y^{(6)}_{\mathbf{1'}}$ ~&~ $Y^{(6)}_{\mathbf{2}}$ ~ 
         \\
         \hline
       ~$S_3$~ & $\mathbf{2}$ & $\mathbf{1}$ & $\mathbf{2}$ & $\mathbf{1}$& $\mathbf{1'}$&$\mathbf{2}$ 
       \\
         \hline
         $-k$~ & 2 & 4 & 4 & 6 & 6 & 6 
         \\
          \hline
    \end{tabular}
    \caption{\centering Modular transformation of Yukawa couplings and their weights.}
    \label{tab:modyuk}
\end{table}

The charged lepton mass matrix derived from the Lagrangian in Eq.~\eqref{eq:yuklag} can be expressed as
\begin{align} \label{eq:clmat}
    m_{\ell}=\frac{v_H}{\sqrt{2}}\begin{pmatrix}
        \alpha_{\ell} Y^{(4)}_{\mathbf{1}} & \beta_{\ell} y_1& \beta_{\ell} y_2 \\
        \gamma_{\ell} Y^{(4)}_{\mathbf{2},1}&- \delta_{\ell} y_1 & \delta_{\ell} y_2 \\
       \gamma_{\ell} Y^{(4)}_{\mathbf{2},2} & \delta_{\ell} y_2& \delta_{\ell} y_1
    \end{pmatrix},
\end{align}
where weight 2 modular Yukawa coupling $Y^{(2)}_{\mathbf{2}} \equiv \left(Y^{(2)}_{\mathbf{2},1}, \ Y^{(2)}_{\mathbf{2},2}  \right)^T \equiv \left( y_1, \ y_2 \right)^T$. 
Then, $m_\ell$ is diagonalized by a bi-unitary mixing matrix as $D_\ell\equiv V_{e_L}^\dag m_\ell V_{e_R}$,
therefore, $|D_\ell|^2 = V_{e_L}^\dag m_\ell m_\ell^\dag V_{e_L}$.
We denote the mass eigenstates of vector-like fermions as $\Psi_k$ and their masses as $M_k$ $(k=1,2)$, given by
\begin{align} \label{eq:fermionmass}
M_1 = \alpha_{\psi} m_0,
\quad  M_2 = \delta_{\psi} m_0, \quad  \text{where} \  m_0\equiv  Y^{(4)}_{\mathbf{1}} \frac{v_{\sigma}}{\sqrt{2}} \equiv  Y^{(4)}_{\mathbf{1}} f_{\rm PQ} ,
\end{align}
where $f_{\rm PQ}$ is the PQ breaking scale. Thus, the PQ breaking scale governs the masses of $\Psi_k$, thereby directly connecting the origin of neutrino masses to the PQ scale.

The neutrino masses are generated at the one loop-level, mediated by colored fermions and scalars, as shown in Fig.~\ref{fig:numass_loop}. 
\begin{figure}[!h]
    \centering
    \begin{tikzpicture}
\begin{feynman}
\vertex (a1) {\(L \)};
\vertex[right=2.0cm of a1] (a2);
\vertex[below=0.05cm of a2] (a3) ;
\vertex[right=2.0cm of a2] (g22);
\vertex[right=0.2cm of g22] (g33);
\vertex[right=2.0cm of g33] (g2) ;
\vertex[below=0.05cm of g2] (g4) ;
\vertex[right=2.0cm of g2] (g3) {\(\nu_{R} \)} ;
\vertex[above=2.1cm of g22] (e1) ;
\vertex[above=1.5cm of e1] (e2) ;
\vertex[below=0.05cm of e1] (e22) ;
\vertex[above=0.05cm of e2] (e5) {\(H\)};
\vertex[below=1.5cm of g22] (f1) {\(\sigma \)};

\diagram* {
(a1) -- [fermion, black,very thick] (a2) -- [fermion,black,very thick, edge label'=\(\Psi_{R} \)] (g22),
(g22) -- [fermion,black,very thick, edge label'=\(\Psi_{L} \)] (g2),
(g2) -- [fermion, black,very thick, edge label' =\hspace{1cm} ] (g3), 
(a2) -- [anti charged scalar,black,very thick, quarter left,edge label=\( \eta \)] (e1) -- [anti charged scalar,black,very thick,quarter left,edge label=\(\chi \)] (g2),
(e1) -- [anti charged scalar,black,very thick] (e2),
(g22) -- [anti charged scalar,black,very thick] (f1),
};
\end{feynman}
\end{tikzpicture}
    \caption{\centering One loop Dirac neutrino masses mediated by colored fields.}
    \label{fig:numass_loop}
\end{figure}
The resulting one loop neutrino mass matrix can be computed as follows
\begin{align} \label{eq:numass}
    \left(m_{\nu} \right)_{\alpha \beta} = \frac{N_c \sin{\theta} \cos{\theta}}{16 \pi^2} \sum_{k=1,2} \mathcal{Y}_{\alpha k} \mathcal{Y}'_{k \beta} M_{k} \left( \frac{m_1^2}{m_1^2-M_{k}^2} \ln{\frac{m_1^2}{M_{k}^2}}-\frac{m_2^2}{m_2^2-M_{k}^2} \ln{\frac{m_2^2}{M_{k}^2}} \right),
\end{align}
where $N_c$ is the color factor, and the angle $\theta$ and masses $m_1$, $m_2$ can be followed from Eqs. \eqref{eq:scalarrot} and \eqref{eq:scalarmass}. The fermion masses $M_{k}$ are given in Eq.~\eqref{eq:fermionmass}. The Yukawa structure is encoded within the matrices $\mathcal{Y}$ and $\mathcal{Y}'$, which are given by
\begin{align} \label{eq:DirYuk}
    \mathcal{Y} \equiv \begin{pmatrix}
        \alpha_{\nu} Y^{(6)}_{\mathbf{1}} &  \beta_{\nu} Y^{(6)}_{\mathbf{1'}} \\
         \gamma_{\nu} Y^{(6)}_{\mathbf{2},1} &  -\delta_{\nu} Y^{(6)}_{\mathbf{2},2} \\
          \gamma_{\nu} Y^{(6)}_{\mathbf{2},2} &  \delta_{\nu} Y^{(6)}_{\mathbf{2},1}
    \end{pmatrix}, \quad
     \mathcal{Y}' \equiv \begin{pmatrix}
        \alpha'_{\nu} Y^{(6)}_{\mathbf{1}} &  \beta'_{\nu} Y^{(4)}_{\mathbf{2},1} & \beta'_{\nu} Y^{(4)}_{\mathbf{2},2}  \\
         \gamma'_{\nu} Y^{(6)}_{\mathbf{1'}} &  -\delta'_{\nu} Y^{(4)}_{\mathbf{2},2} & \delta'_{\nu} Y^{(4)}_{\mathbf{2},1}
    \end{pmatrix}.
\end{align}
As discussed earlier, we introduce only two generations of vector-like fermions $\Psi_k$. As a result, the neutrino mass matrix is of rank two, yielding masses for two neutrinos, while the lightest neutrino remains massless. This feature has important implications for the model, particularly for the sum of neutrino masses, as explored in the numerical analysis presented in the subsequent section.

In order to analyze the neutrino oscillation, we redefine $m_\nu$ to be $m_\nu\equiv \kappa_\nu \tilde m_\nu$,
where 
\begin{align}
\kappa_\nu\equiv 
 \frac{N_c \sin{\theta} \cos{\theta}}{16 \pi^2} \alpha_\nu \alpha'_\nu m_0 ,
\end{align}
here $m_0\equiv  Y^{(4)}_{\mathbf{1}} \frac{v_{\sigma}}{\sqrt{2}}\equiv Y^{(4)}_{\mathbf{1}} f_{\rm PQ}$.
Notably, neutrino masses are intimately connected to the PQ breaking scale $f_{\rm PQ}$, which in turn directly relates them to the axion mass.
The neutrino mass matrix $m_\nu$ is diagonalized by a bi-unitary mixing matrix as
$D_\nu\equiv V_{\nu_L}^\dag m_\nu V_{\nu_R} = \kappa_\nu V_{\nu_L}^\dag \tilde m_\nu V_{\nu_R}$, therefore, $|\tilde D_\nu|^2 = V_{\nu_L}^\dag \tilde m_\nu\tilde  m_\nu^\dag V_{\nu_L}$. Then, we can rewrite $\kappa_\nu$ in terms of atmospheric mass-squared difference $\Delta m^2_{\rm atm}$ and mass eigenvalues of neutrinos as follows:
\begin{align}
({\rm NH}):\ \kappa_\nu^2= \frac{\Delta m^2_{\rm atm}}{|\tilde D_{\nu_3}|^2},\\
({\rm IH}):\ \kappa_\nu^2= \frac{\Delta m^2_{\rm atm}}{|\tilde D_{\nu_2}|^2}.
\end{align}
The solar mass-squared difference $\Delta m^2_{\rm sol}$ is given by
\begin{align}
({\rm NH}):& \ \Delta m^2_{\rm sol} =\Delta m^2_{\rm atm} 
\left(\frac{|\tilde D_{\nu_2}|^2}{|\tilde D_{\nu_3}|^2}\right),\\
({\rm IH}):& \ \Delta m^2_{\rm sol} =\Delta m^2_{\rm atm} 
\left(\frac{|\tilde D_{\nu_2}|^2-|\tilde D_{\nu_1}|^2}{|\tilde D_{\nu_2}|^2}\right).
\end{align}
Through these relations, the neutrino mass eigenvalues are found as 
\begin{align}
({\rm NH}):& \ D_{\nu_2}= \sqrt{\Delta m^2_{\rm sol}} ,\ 
D_{\nu_3}= \sqrt{\Delta m^2_{\rm atm}} \ , \\
({\rm IH}):& \ D_{\nu_1}= \sqrt{\Delta m^2_{\rm atm} - \Delta m^2_{\rm sol}} ,\ 
D_{\nu_2}= \sqrt{\Delta m^2_{\rm atm}} \ .
\end{align}
The observed mixing matrix for the lepton sector is given by $U_{\rm PMNS}\equiv V_{e_L}^\dag V_{\nu_L}$.
Applying the standard parametrization in PDG~\cite{ParticleDataGroup:2018ovx}, we find three mixing angles and Dirac CP phase; $\delta_{\rm CP}$, to be
\begin{align}
s_{13}= |U_{\rm PMNS}|_{13},\quad 
s_{12}= |U_{\rm PMNS}|_{12}/c_{13},\quad 
s_{23}= |U_{\rm PMNS}|_{23}/c_{13},\quad 
\delta_{\rm CP} = {\rm arg}[(U_{\rm PMNS})_{13}],
\end{align}
where $s_{13,12,23}$ ($c_{13,12,23}$) is abbreviation form of $\sin\theta_{13,12,23}$ ($\cos\theta_{13,12,23}$) in the three mixing angles.

\section{Numerical analysis and predictions} \label{sec:numericalanalysis}

In this section, we present our numerical $\chi^2$ analysis to satisfy the neutrino oscillation data.
At first, we randomly select our input parameters of the neutrino sector in the following ranges:
\begin{align}
[\beta^{(')}_\nu/\alpha^{(')}_\nu,\gamma^{(')}_\nu/\alpha^{(')}_\nu,
|\delta^{(')}_\nu/\alpha^{(')}_\nu|, m_i/m_0, M_i/m_0] : [10^{-5},10^5],\quad
\arg\left(\frac{\delta^{(')}_\nu}{\alpha^{(')}_\nu}\right) : [0,2\pi].
\end{align}
Here, \(\beta^{(')}_\nu/\alpha^{(')}_\nu\) and
\(\gamma^{(')}_\nu/\alpha^{(')}_\nu\) are treated as real parameters,
with both signs allowed, whereas
\(\delta^{(')}_\nu/\alpha^{(')}_\nu\) is treated as a complex
parameter. We define
\(m_0\equiv Y^{(4)}_{\mathbf{1}}v_{\sigma}/\sqrt{2}\) and take
\(m_{\eta^\pm}\simeq m_1\) to satisfy the constraints from the oblique
parameters~\cite{Barbieri:2006dq}.
In the scan, we additionally require perturbative couplings, a bounded from below scalar potential, and positive physical scalar masses to avoid tachyonic directions.
In addition to the parameters of the neutrino sector, the charged lepton sector contains free parameters of its own.
Without loss of generality, we restrict $\tau$ to the fundamental domain of the modular group.
For a given $\tau$, the charged lepton mass matrix in Eq.~\eqref{eq:clmat} depends on four real parameters, $\alpha_\ell$, $\beta_\ell$, $\gamma_\ell$, and $\delta_\ell$, which are subject to three constraints from the observed charged lepton masses. One degree of freedom therefore remains, which we take to be $\alpha_\ell$.
For each sampled value of $\tau$, the ratios $\beta_\ell/\alpha_\ell$, $\gamma_\ell/\alpha_\ell$, and $\delta_\ell/\alpha_\ell$ are then fixed by the invariant relations
\begin{align}
&{\rm Tr}[m^\dag_\ell m_\ell] = |m_e|^2 + |m_\mu|^2 + |m_\tau|^2,\\
& {\rm Det}[m^\dag_\ell m_\ell] = |m_e|^2  |m_\mu|^2  |m_\tau|^2, \\
&\frac{({\rm Tr}[m^\dag_\ell m_\ell])^2 -{\rm Tr}[(m^\dag_\ell m_\ell)^2]}2 = |m_e|^2  |m_\mu|^2 + |m_\mu|^2  |m_\tau|^2+ |m_e|^2  |m_\tau|^2.
\label{eq:l-cond0}
\end{align}
In our numerical analysis we fix $\alpha_\ell$ to the representative value $\alpha_\ell \simeq 0.01$.%
\footnote{Varying $\alpha_\ell$ over $[10^{-5},1]$ only mildly shifts the allowed range of $\mathrm{Im}\,[\tau]$ and leaves that of $\mathrm{Re}\,[\tau]$ unchanged.
Compared with the fixed $\alpha_\ell$ case, its width increases by a factor of about $1.3$ for NH and about $5.9$ for IH.
Hence, for NH the results obtained with fixed $\alpha_\ell$ are representative, while for IH they should be regarded as a conservative subset of the parameter space allowed when $\alpha_\ell$ is free.}
This is simply a choice made to keep the parameter space of the scan minimal and does not imply that $\alpha_\ell$ is fixed by the charged lepton mass constraints. Rather, it corresponds to fixing the one remaining degree of freedom after imposing the three charged lepton mass constraints.
Then, we make use of five reliable observables provided by NuFIT 6.0~\cite{Esteban:2024eli}: $[\Delta m^2_{\rm atm},\Delta m^2_{\rm sol}, s^2_{12}, s^2_{13}, s^2_{23}]$ for our analysis, and we show our allowed region within the 5$\sigma$ confidence level.

In the minimal model, the rank two neutrino mass matrix leaves one neutrino massless. Consequently, once the measured mass-squared differences are imposed, both the sum of neutrino masses $\sum D_{\nu}$ and the effective
electron-neutrino mass $m_\beta$ are restricted to narrow ranges. These ranges are generic consequences of a one massless neutrino spectrum rather than distinctive predictions of the modular $S_3$ symmetry or the axion sector.
We note that, within the allowed parameter space, we find no significant correlations among the mixing angles $\theta_{12}$, $\theta_{13}$, $\theta_{23}$ and the Dirac CP-violating phase $\delta_{\rm CP}$. This is a consequence of the sufficient freedom in the model parameters, which prevents the modular $S_3$ structure from yielding additional predictive correlations among these observables.
The corresponding figures presented below are retained to illustrate these phenomenological consequences and to compare them with current experimental and cosmological constraints.

\subsection{Normal Hierarchy (NH)}
\begin{figure}[!h]
    \centering
    \includegraphics[width=0.55\linewidth]{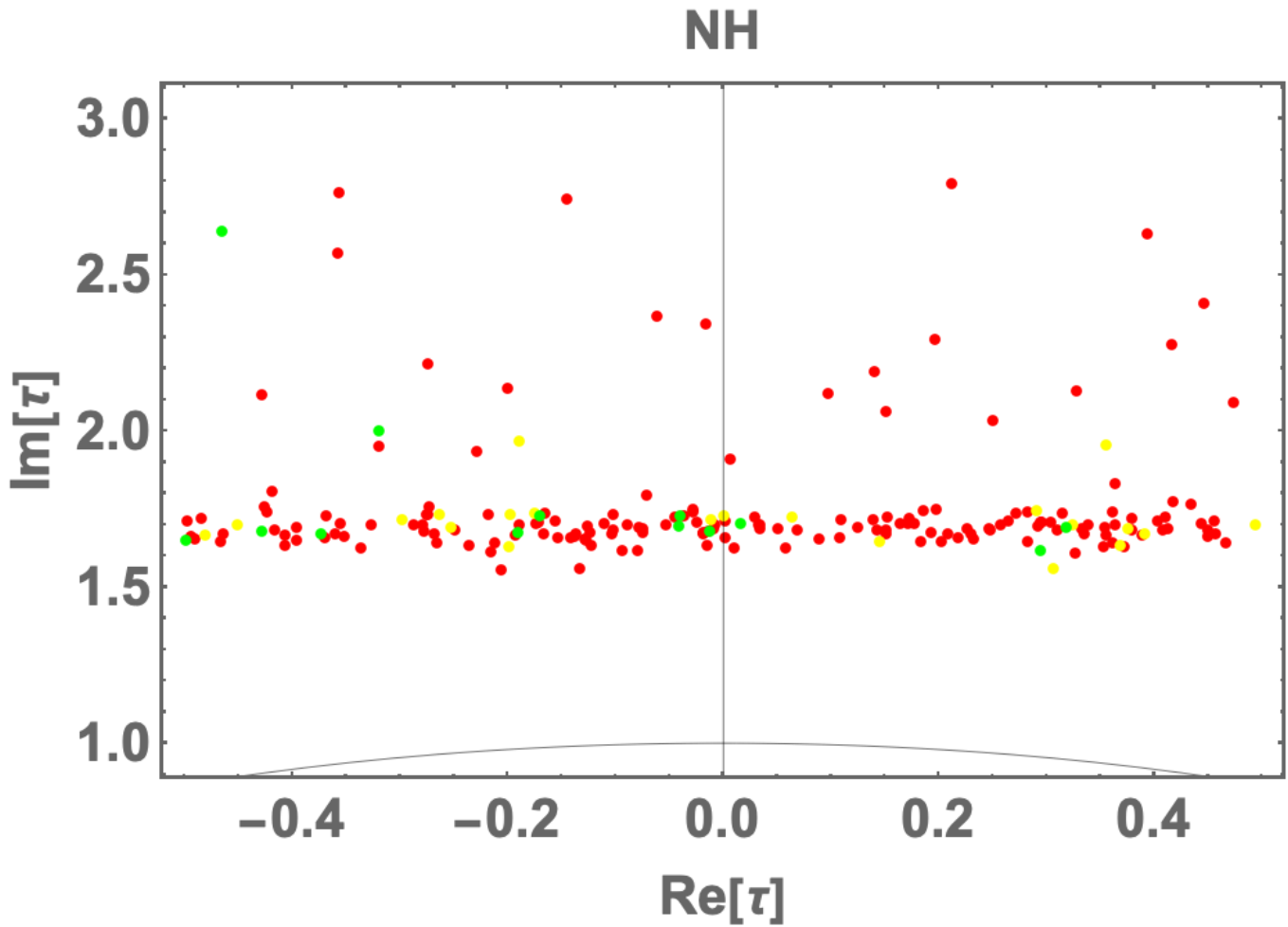}
    \caption{Allowed region of Re[$\tau$] and Im[$\tau$] on the fundamental region in case of NH keeping $\alpha_\ell$ fixed. Here, green points represent the range of ($1\sigma-2\sigma$), yellow ones ($2\sigma-3\sigma$), and red ones ($3\sigma-5\sigma$).}
    \label{fig:tau_nh}
\end{figure}

We begin with a discussion of the NH case. From a $\chi^2$ analysis, the best-fit (BF) values of the parameters are obtained at $\chi_{\rm min} =2.48$. These values are summarized in Table \ref{tab:BFNH}.
\begin{table}[!h]
    \setlength\tabcolsep{0.2cm}
    \begin{tabular}{c|c||c|c||c|c}
\hline
        Parameter  &  BF & Parameter & BF & Parameter & BF \\ \hline \hline
          $\tau$ &  $-0.0411+1.69 i $ & $m_{1}/m_0$ & $2.13\times10^3$ & $M_{1}/m_0$ &$6.73\times10^{-5}$ \\ \hline
       $\beta^{}_\nu/\alpha^{}_\nu$  &$2.12\times10^{-4}$ &  $\gamma^{}_\nu/\alpha^{}_\nu$  &$1.27$ &
 $M_{2}/m_0$ & $908$ \\ \hline
 $\beta^{'}_\nu/\alpha^{'}_\nu$  &$0.0158$ &  $\gamma^{'}_\nu/\alpha^{'}_\nu$  &$-1.75\times10^{-4}$ &
   &   \\ \hline
 $\delta_\nu/\alpha_\nu$ ($ 10^{-5}$)  &$(-1.27 + 1.31i)$ 
 & $\delta^{'}_\nu/\alpha^{'}_\nu$ ($10^{-4}$) &$(1.75 + 6.63 i)$ 
 & $m_{2}/m_0$ & $2.47\times10^3$ 
 \\ \hline
$\beta_{\ell}/\alpha_\ell$ & $0.993$ & $\delta_{\ell}/\alpha_\ell$ 
& $0.000365$ & $\gamma_{\ell}/\alpha_\ell$ & $0.123$ \\ \hline
$s_{12}$ & 0.569 &$s_{23}$ & 0.734 &$s_{13}$ & 0.148 \\ \hline
$\Delta m^2_{\rm sol}$ & $7.50\times 10^{-5}~{\rm eV}^2$ &  $\Delta m^2_{\rm atm}$ & $2.51\times 10^{-3}~{\rm eV}^2$ &
$\delta_{\rm CP}$ & $330^{\circ}$ \\ \hline
$\sum D_{\nu}$ & $58.8~{\rm meV}$ & $m_{\beta}$ 
& $9.03~ {\rm meV}$ & $|\kappa_ \nu|^2$ & $4.49\times 10^{-11}$  
\\
\hline
    \end{tabular}
    \caption{\label{tab:BFNH}%
      Best-fit (BF) parameter values in the NH case corresponding to $\chi_{\rm min} =2.48$.}
\end{table}
We present our model predictions through the scattered plots. In Fig.~\ref{fig:tau_nh}, we show the allowed region on Im[$\tau$] in terms of Re[$\tau$], where the blue points represent the range $\le1\sigma$, green ones ($1\sigma-2\sigma$), yellow ones ($2\sigma-3\sigma$), and red ones ($3\sigma-5\sigma$)\footnote{In the case of NH, we have null results within the range $\le1\sigma$; blue points.}.
The figure tells us that the whole range of Re[$\tau$] is allowed, but Im[$\tau$] is restricted to the range [1.56-2.79].
Moreover, the allowed region is localized at Im[$\tau$]$\sim1.7$.
When $\alpha_\ell$ is allowed to vary, the corresponding allowed range of $\mathrm{Im}[\tau]$ becomes [1.31-2.95].

\begin{figure}[!h]
    \centering
    \includegraphics[width=0.48\linewidth]{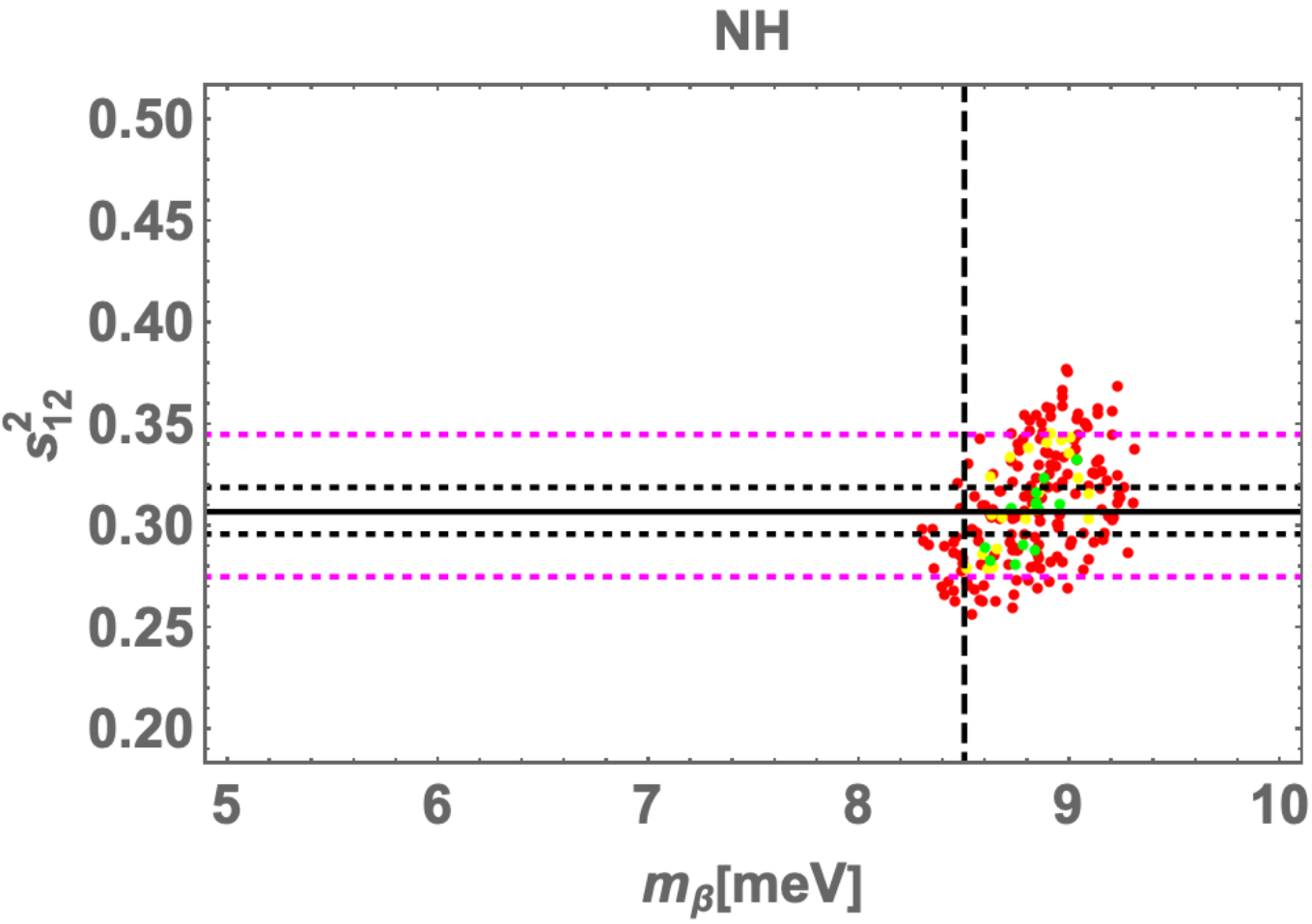}
    \hspace{0.1cm}
    \includegraphics[width=0.495\linewidth]{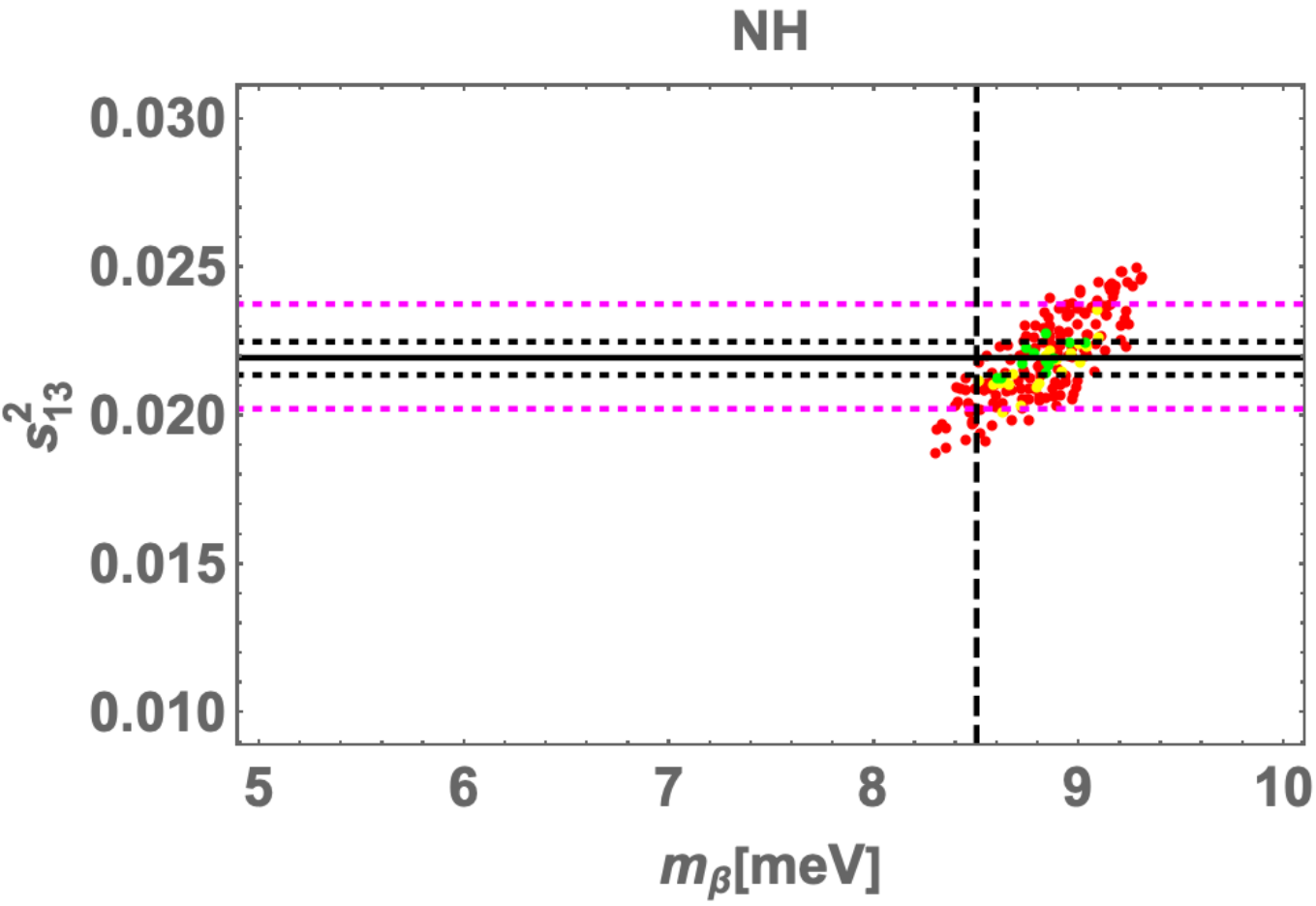}
    \caption{Allowed regions of $s^2_{12}$ (left), $s^2_{13}$ (right) in terms of $m_\beta$, where the color legends of the plots are the same as in Fig.~\ref{fig:tau_nh}.}
    \label{fig:mnue_angles_nh}
\end{figure}

In Fig.~\ref{fig:mnue_angles_nh}, we show the allowed regions of $s^2_{12}$ (left) and $s^2_{13}$ (right) in terms of the effective electron-neutrino mass $m_{\beta}$, expressed in units of meV. The color legends of the plots are the same as in Fig. \ref{fig:tau_nh}.
The effective mass $m_{\beta}$ is defined as
\begin{align} \label{eq:me}
 | m_{\beta}| \equiv \sqrt {\sum_{i}  |U_{ei} |^2 D^2_{\nu_i}}  = \sqrt{c^2_{12}c^2_{13}D^2_{\nu_1} + s^2_{12}c^2_{13} D^2_{\nu_2} +s^2_{13}D^2_{\nu_3}  }\;.
\end{align}
This expression is independent of the mixing angle $\theta_{23}$. Since one neutrino is massless in the rank two spectrum, the allowed
range of $m_{\beta}$ is largely determined by the measured mass-squared differences and mixing angles. Fig.~\ref{fig:mnue_angles_nh}
therefore illustrates the resulting phenomenological ranges and their
comparison with current experimental constraints.
The horizontal lines represent the best-fit value (black line), the 1$\sigma$ interval (dotted black), and the 3$\sigma$ interval (dotted magenta) from two variants $\chi^2$ analysis in Table 1 of NuFIT 6.0.

\begin{figure} [!h]
    \centering
    \includegraphics[width=0.495\linewidth]{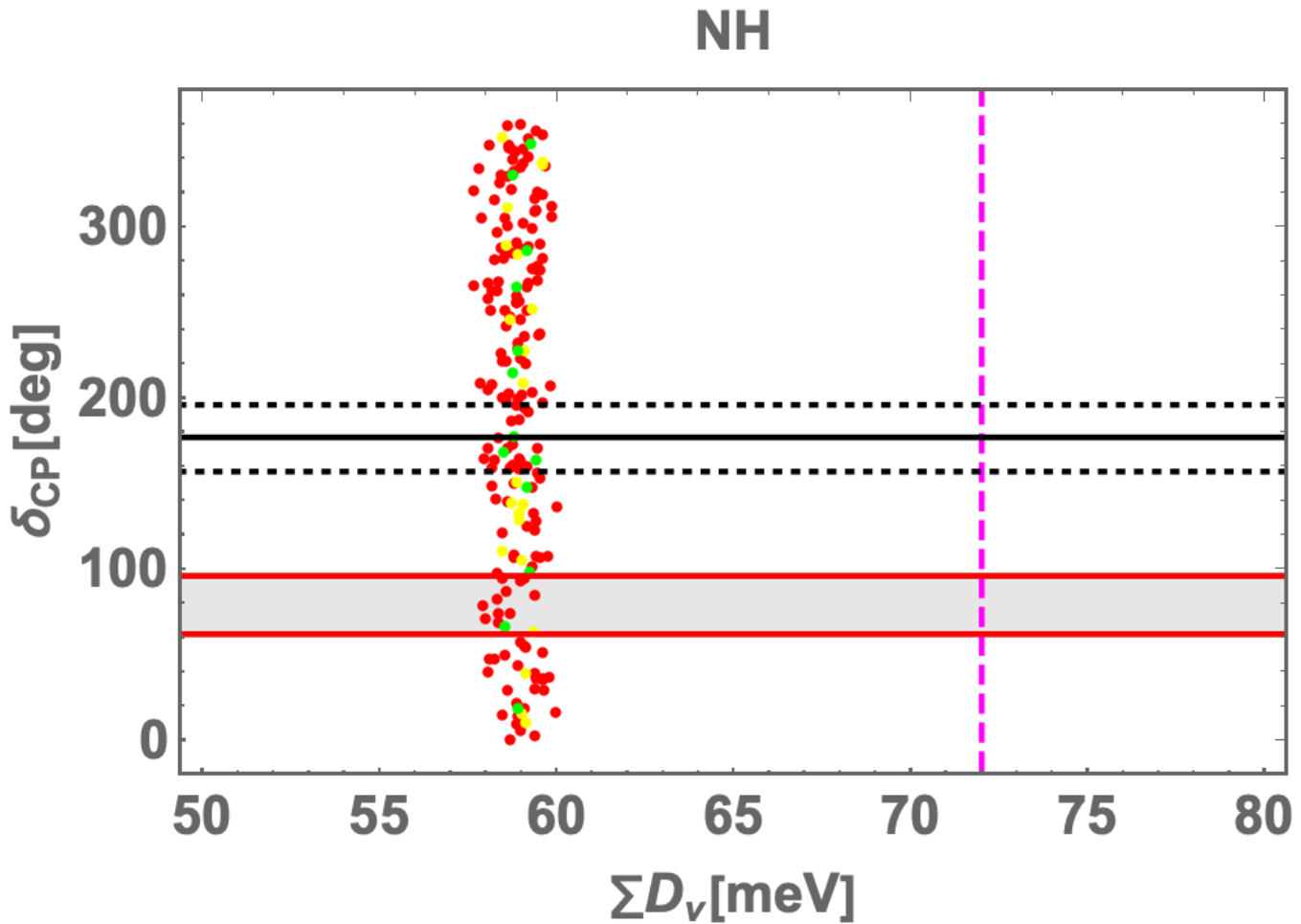}
    \hspace{0.1cm}
    \includegraphics[width=0.48\linewidth]{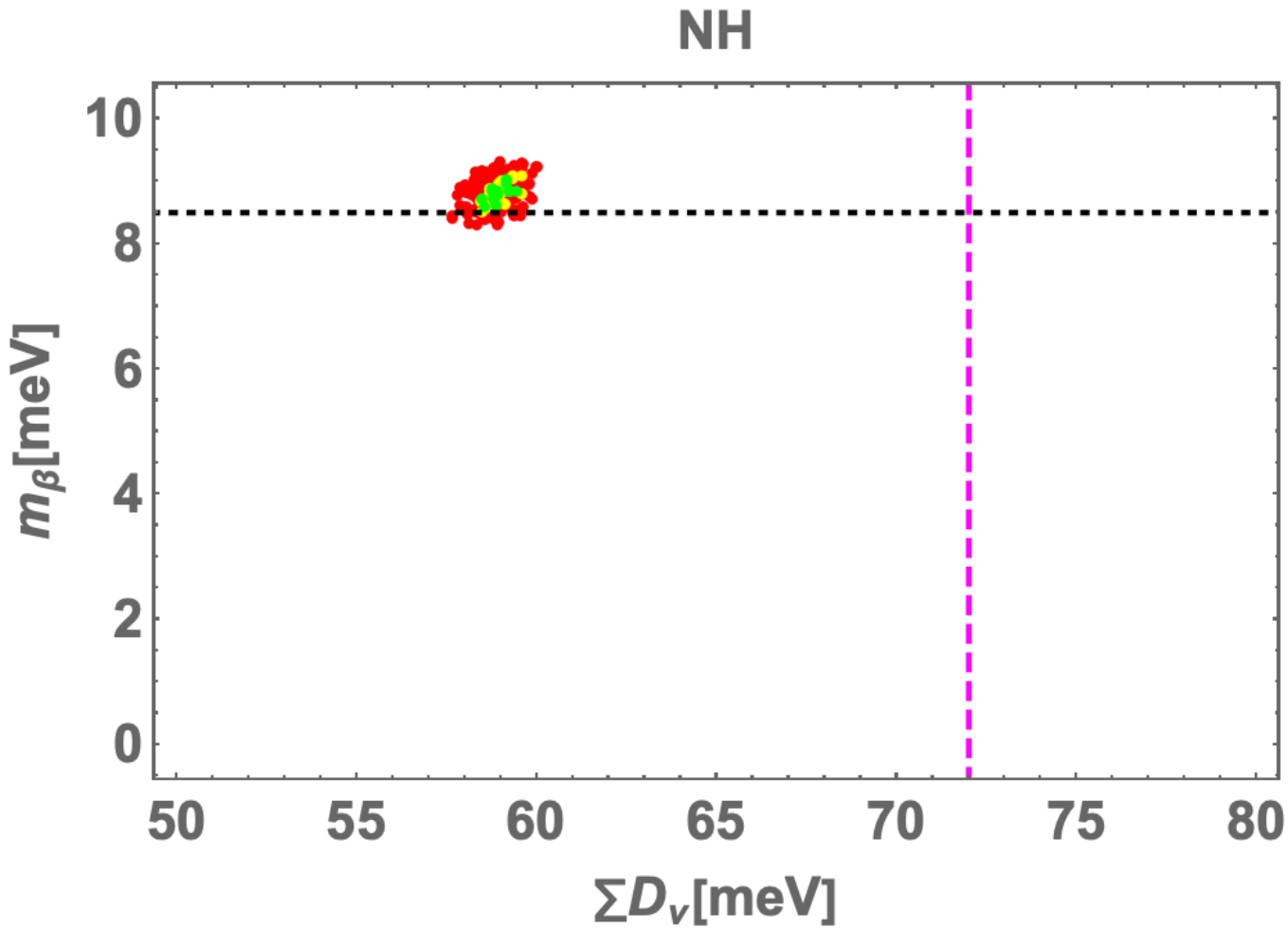}
    \caption{Allowed regions of $\delta_{\rm CP}$ (left) and $m_{\beta}$ (right) in terms of $\sum D_\nu$, where the color legends of the plots are the same as in Fig.~\ref{fig:tau_nh}.}
    \label{fig:sum_nh}
\end{figure}

Fig.~\ref{fig:sum_nh} shows the allowed regions of the Dirac CP phase $\delta_{\rm CP}$ (left) and $m_{\beta}$ (right) in terms of the sum of neutrino masses $\sum D_\nu$ in units of meV. 
The color legends of the plots are the same as in Fig. \ref{fig:tau_nh}. The dashed magenta vertical line at $\sum D_\nu=72$ meV shows the upper bound from the recent combined data of DESI and CMB~\cite{DESI:2024mwx}.
The sum of neutrino masses is restricted to $[57.6,60.0]$ meV, as expected from the rank two neutrino spectrum with one massless neutrino and the measured values of $\Delta m^2_{\rm atm}$ and $\Delta m^2_{\rm sol}$. The cosmological bound shown in
Fig.~\ref{fig:sum_nh} is included only for comparison. Its implications for a particular mass ordering with $m_{\rm lightest}=0$ apply generally to the corresponding neutrino mass spectrum within the cosmological assumptions underlying this bound.
In the left panel, the solid black horizontal line at $177^\circ$ shows the BF value, while the dotted black lines at
$177^{+19}_{-20}{}^\circ$ indicate the $1\sigma$ interval. The gray
region, $(62^\circ\!-\!96^\circ)$, is excluded at $3\sigma$ by NuFIT~6.0.
In the right panel, the dotted black horizontal line at $m_{\beta}=8.5$ meV marks the lower end of the 95\% confidence interval
obtained from the global oscillation analysis together with the KATRIN
bound~\cite{Esteban:2024eli,KATRIN:2024cdt}
\footnote{
The KATRIN experiment gives the upper limit
$m_{\beta}<450$ meV at the 90\% confidence level~\cite{KATRIN:2024cdt}. The global oscillation analysis combined with the KATRIN result gives
$8.5~\mathrm{meV}\leq m_{\beta}\leq400~\mathrm{meV}$
at the 95\% confidence level for the NH scenario~\cite{Esteban:2024eli}.}.
The figure shows that the Dirac CP phase spans nearly the full range, while $m_\beta$ is restricted to
$[8.30,9.31]$ meV.
The nearly vertical bands in Fig.~\ref{fig:sum_nh} therefore reflect
the narrow allowed range of $\sum D_\nu$ and illustrate the generic
phenomenological consequence of a neutrino spectrum with one massless
state.
\subsection{Inverted Hierarchy (IH)}
\begin{figure}[!h]
    \centering
    \includegraphics[width=0.55\linewidth]{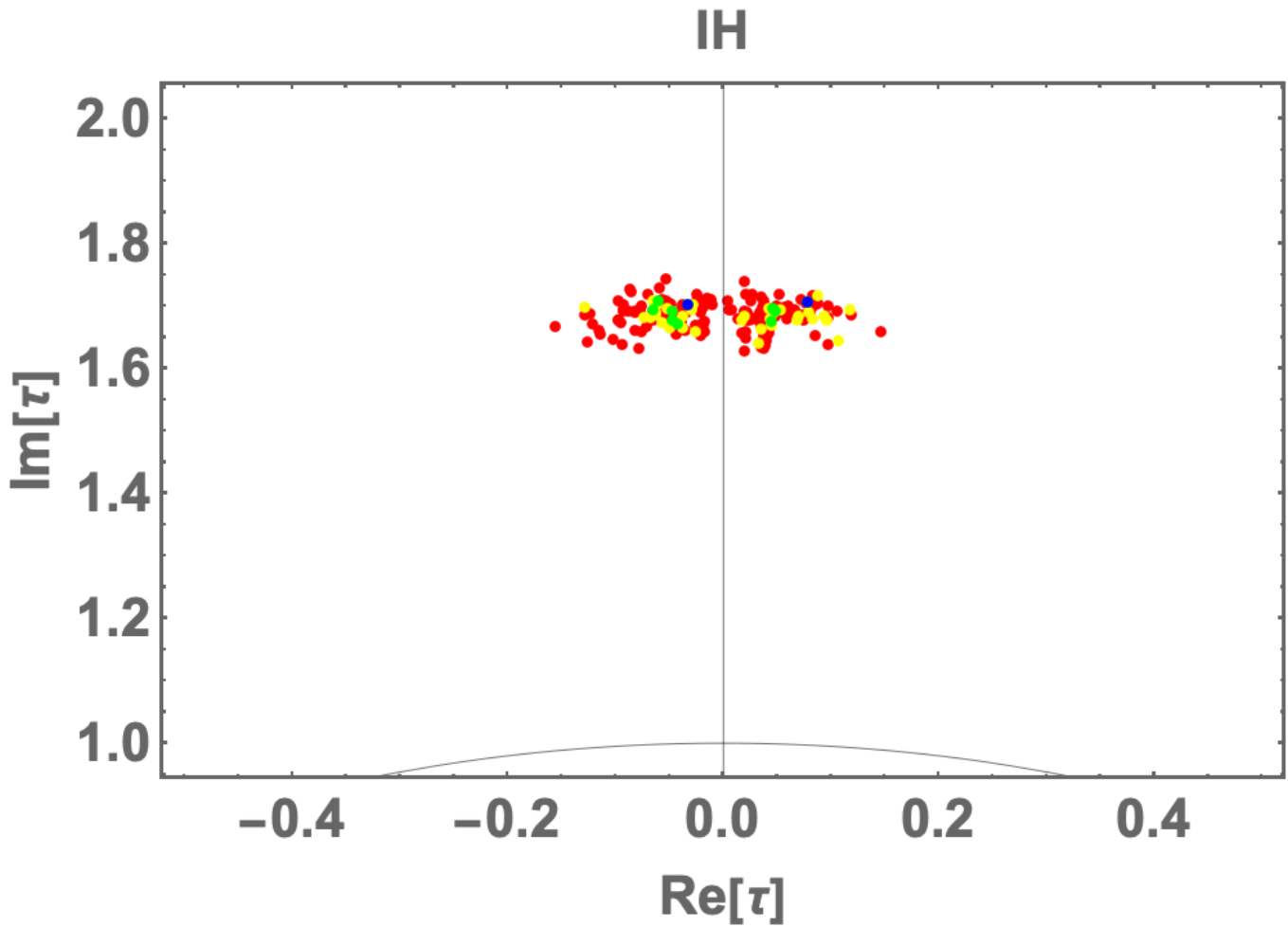}
    \caption{Allowed region of Im[$\tau$] in terms of Re[$\tau$] keeping $\alpha_\ell$ fixed, where the color legends of the plots are the same as in Fig.~\ref{fig:tau_nh}.}
    \label{fig:tau_ih}
\end{figure}

\begin{figure}[!h]
    \centering
    \includegraphics[width=0.48\linewidth]{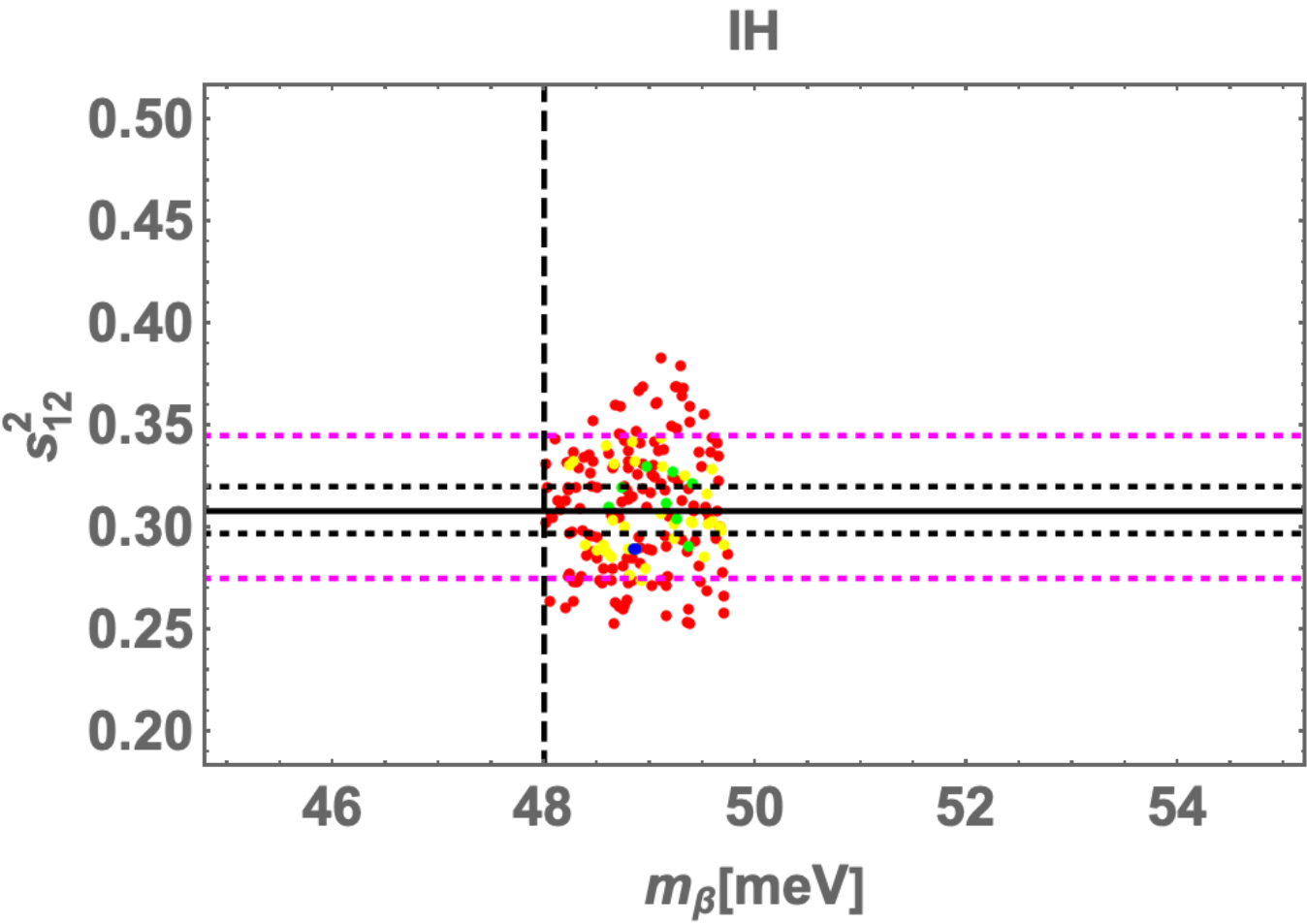}
    \hspace{0.1cm}
    \includegraphics[width=0.495\linewidth]{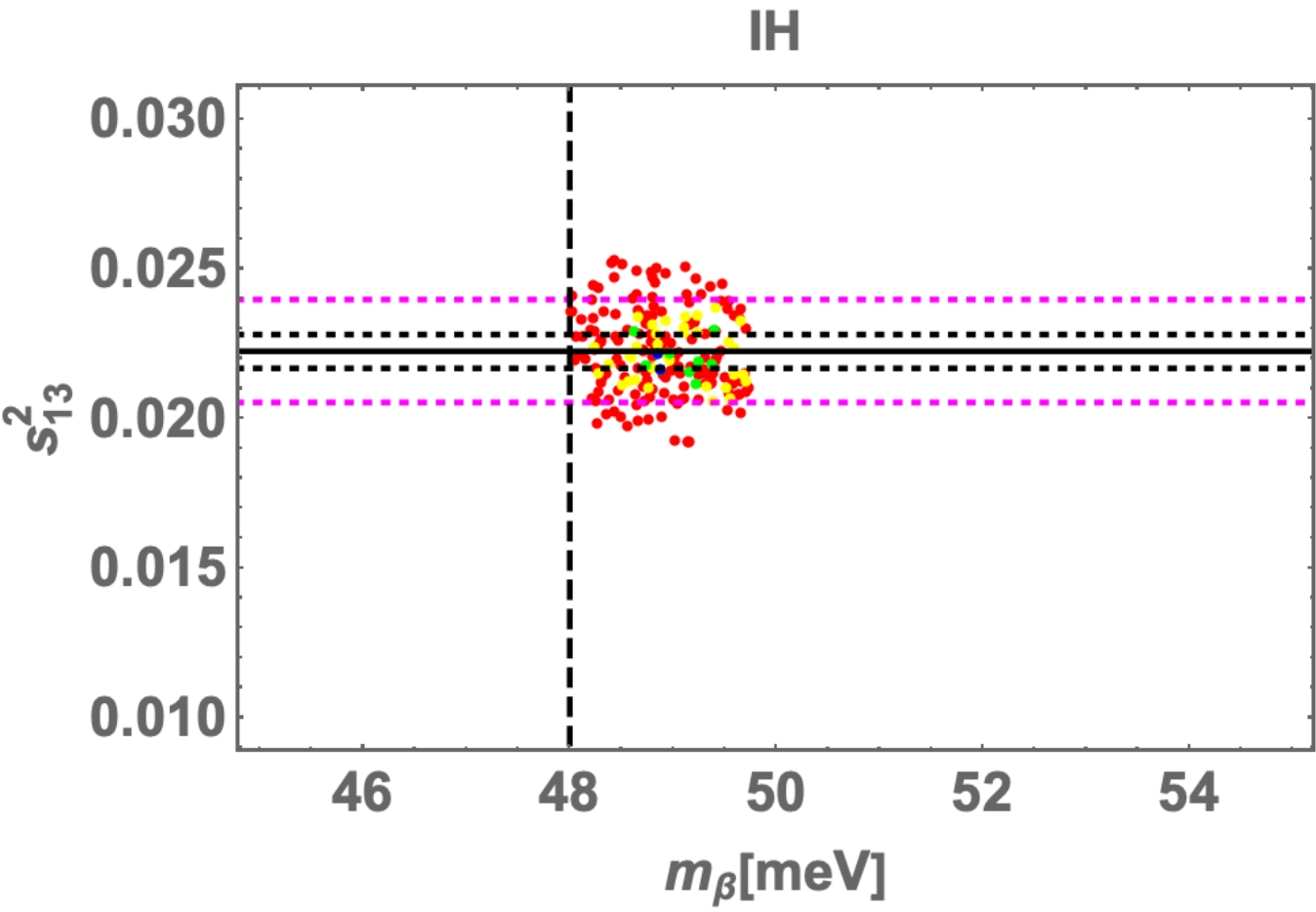}
    \caption{Allowed regions of $s^2_{12}$ (left), $s^2_{13}$ (right) in terms of $m_\beta$, where the color legends of the plots are the same as in Fig.~\ref{fig:tau_nh}.}
    \label{fig:mnue_angles_ih}
\end{figure}

We now proceed to discuss the IH scenario. For this case, the BF values of the parameters are obtained at $\chi_{\rm min} =1.88$ and summarized in Table \ref{tab:BFIH}.
\begin{table}[!h]
    \setlength\tabcolsep{0.2cm}
    \begin{tabular}{c|c||c|c||c|c}
\hline
        Parameter    &  BF & Parameter & BF & Parameter & BF \\ \hline \hline
          $\tau$ &  $-0.0332+1.70 i $ & $m_{1}/m_0$ & $6.01\times10^3$ & $M_{1}/m_0$ &$1.59\times10^{3}$ \\ \hline
       $\beta^{}_\nu/\alpha^{}_\nu$  &$-2.39\times10^{-2}$ &  $\gamma^{}_\nu/\alpha^{}_\nu$  &$-0.790$ &
 $M_{2}/m_0$ & $122$ \\ \hline
$\beta^{'}_\nu/\alpha^{'}_\nu$  &$-169$ &  $\gamma^{'}_\nu/\alpha^{'}_\nu$  &$35.5$ &
 $ $ & $ $ \\ \hline
 $\delta_\nu/\alpha_\nu$  &$0.733+0.492i$ &
 $\delta^{'}_\nu/\alpha^{'}_\nu$  &$(1.48 + 2.24 i)\times10^{3}$ & $m_{2}/m_0$ & $5.89\times10^3$ \\ \hline
$\beta_{\ell}/\alpha_\ell$ & $0.994$ & $\delta_{\ell}/\alpha_\ell$ 
& $0.000365$ & $\gamma_{\ell}/\alpha_\ell$ & $0.123$ \\ \hline
$s_{12}$ & 0.538 &$s_{23}$ & 0.742 &$s_{13}$ & 0.149 \\ \hline
$\Delta m^2_{\rm sol}$ & $7.47\times 10^{-5}~{\rm eV}^2$ &  $\Delta m^2_{\rm atm}$ & $2.49\times 10^{-3}~{\rm eV}^2$ &
$\delta_{\rm CP}$ & $332^{\circ}$ \\ \hline
$\sum D_{\nu}$ & $99.1~{\rm meV}$ & $m_{\beta}$ 
& $48.9~ {\rm meV}$ & $|\kappa_ \nu|^2$ & $3.04\times 10^{-28}$ \\
\hline
    \end{tabular}
    \caption{\label{tab:BFIH}%
      Best-fit (BF) parameter values in the IH case corresponding to $\chi_{\rm min} =1.88$.}
\end{table}
In Fig.~\ref{fig:tau_ih}, we show the allowed region of Im[$\tau$] in terms of Re[$\tau$], where the color legends of the plots are the same as in Fig.~\ref{fig:tau_nh}.
The figure tells us that Re[$\tau$] and Im[$\tau$] are respectively localized at $[-0.156,0.147]$ and $[1.63,1.74]$. When $\alpha_\ell$ is allowed to vary, the corresponding allowed range of $\mathrm{Im}[\tau]$ becomes [1.26-1.91].
For the IH case, we similarly show the allowed regions of $s_{12}^2$ (left) and $s_{13}^2$ (right) in terms of
$m_\beta$ in Fig. \ref{fig:mnue_angles_ih}, using the same color coding as in
Fig.~\ref{fig:mnue_angles_nh}.
The vertical dotted black line $(m_{\beta}=48$ meV) corresponds to the lowest upper bound on the global analysis of oscillation data \cite{Esteban:2024eli}\footnote{The latest global analysis of neutrino oscillation data provides $48~\mathrm{meV}\leq m_{\beta}\leq 400~\mathrm{meV}$ for the IH scenario \cite{Esteban:2024eli}.}. However, the KATRIN experiment currently provides an upper bound of $m_{\beta}<450$ meV at the 90\% confidence level, which is substantially larger than $m_{\beta}=48$ meV~\cite{KATRIN:2024cdt}.
The mixing angles span the allowed ranges shown in the figure, while $m_{\beta}$ is restricted to $[48.0,49.7]$ meV. This predicted range starts close to the lower limit of $48$ meV from the global analysis~ \cite{Esteban:2024eli}.

\begin{figure} [!h]
    \centering
    \includegraphics[width=0.49\linewidth]{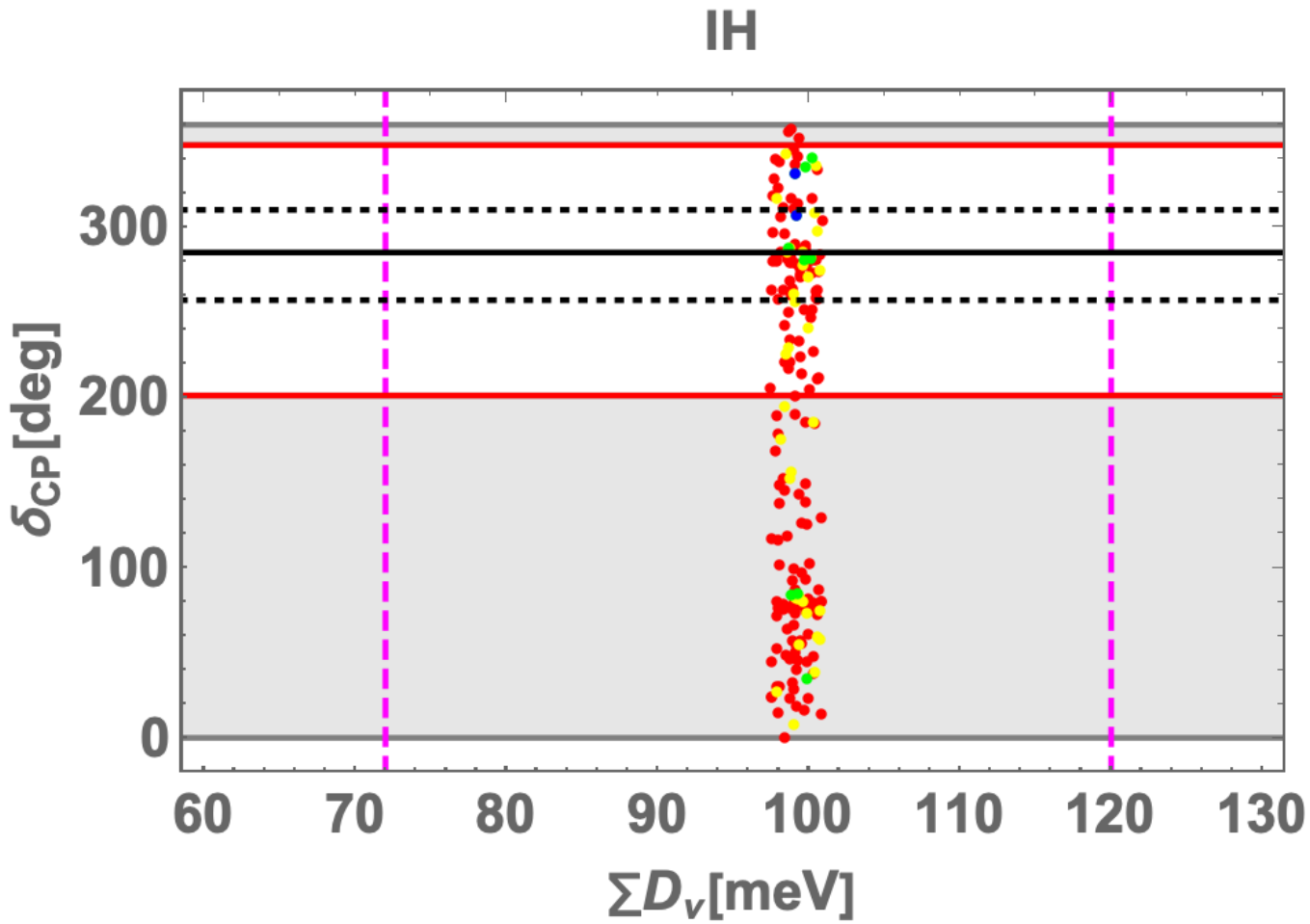}
    \hspace{0.2cm}
    \includegraphics[width=0.48\linewidth]{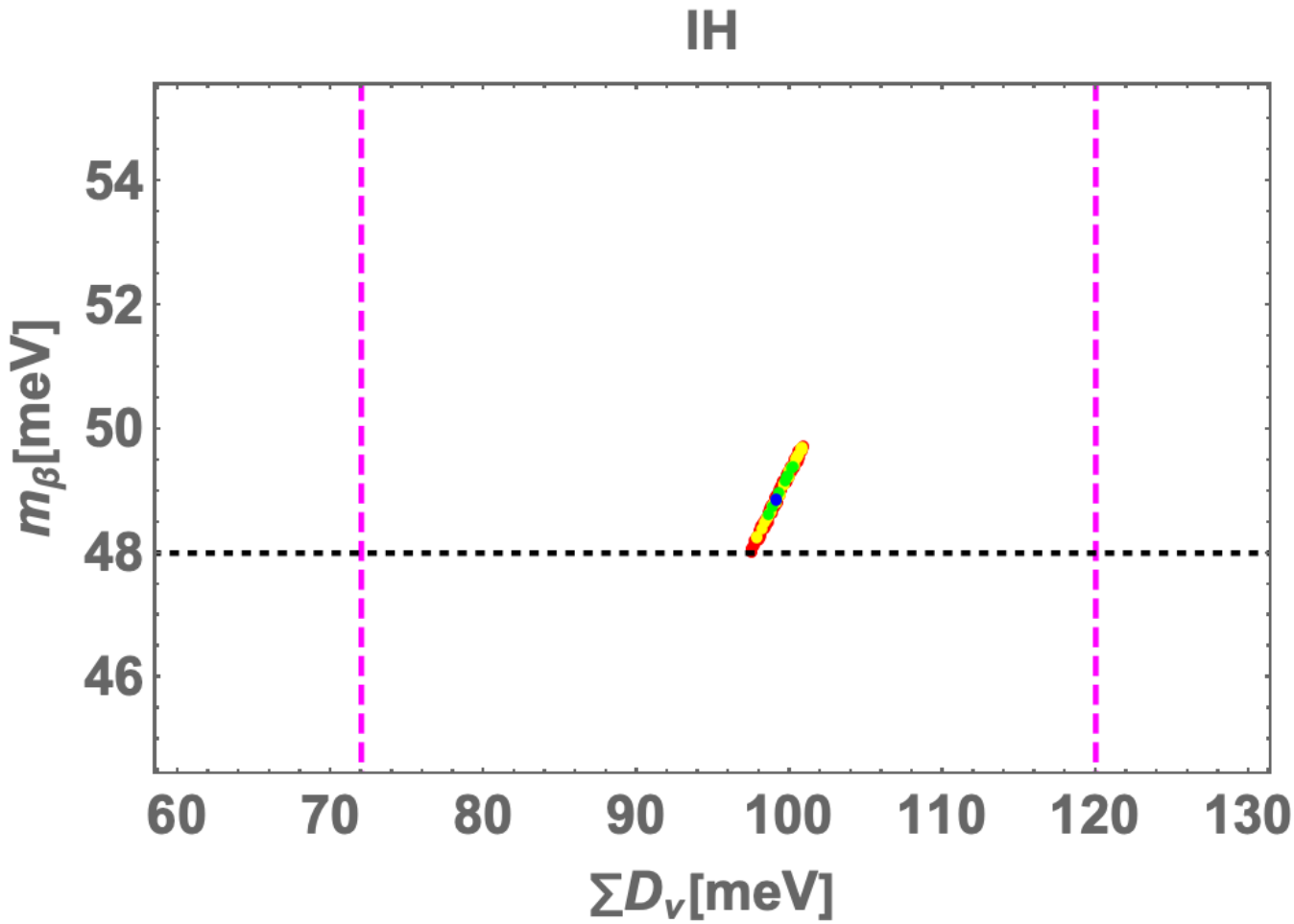}
    \caption{Allowed regions of $\delta_{\rm CP}$ (left) and $m_{\beta}$ (right) in terms of $\sum D_\nu$, where the color legends of the plots and vertical magenta dashed lines are the same as in Fig.~\ref{fig:sum_nh}.}
    \label{fig:sum_ih}
\end{figure}

For the sake of comparison with the NH case, Fig.~\ref{fig:sum_ih} presents the allowed regions of $\delta_{\rm CP}$ (left) and $m_{\beta}$ (right) in terms of $\sum D_\nu$, where the color legends of the plots and the vertical magenta dashed line ($\sim 72$ meV) have the same meanings as in Fig.~\ref{fig:sum_nh}. The additional vertical magenta dashed line denotes the upper bound on $\sum D_\nu$ from the Planck data~\cite{Planck:2018vyg}.
In the left panel, the solid black horizontal line at $285^\circ$  shows the BF value, the dotted black lines show the $1\sigma$ interval $285^{+25}_{-28}{}^\circ$, and the gray regions $[0^\circ,201^\circ]$ and $[348^\circ,360^\circ]$ are excluded at $3\sigma$ by NuFIT~6.0. In the right panel, the dotted black horizontal line at $m_{\beta}=48$ meV marks the lower end of the 95\% confidence interval from the global analysis~ \cite{Esteban:2024eli}.
The right panel illustrates the expected kinematic relation between $m_\beta$ and $\sum D_\nu$, while the left panel forms a nearly vertical band. As in the NH case, the narrow range of $\sum D_\nu$ in both panels is a direct consequence of one massless neutrino and the measured mass-squared differences.
It is evident from the figure that, for the IH case, the predicted range of $\sum D_\nu$ is consistent with the Planck limit~\cite{Planck:2018vyg}, but lies outside the constraint obtained from the combined DESI and CMB data~\cite{DESI:2024mwx}.

\section{Axion and its role as Dark Matter} \label{sec:axion}
\vspace{-0.35cm}
We now delve into a detailed discussion of the axion and its potential role as a DM candidate. 
The complex scalar $\sigma$ can be parameterized as $\sigma = \left (v_{\sigma} + \rho  \right) \exp{(i a/v_{\sigma})}/\sqrt{2}$, where $a$ corresponds to the axion and $\rho$ is the radial mode. Once the scalar $\sigma$ acquires a vev, the $\rm PQ$ symmetry is spontaneously broken at a scale $f_{\rm PQ} \equiv \langle \sigma \rangle = v_{\sigma}/\sqrt{2}$. The axion decay constant $f_a$ is directly related to the $\rm PQ$ breaking scale $f_{\rm PQ}$, which is given by
\begin{align}
    f_a =\frac{f_{\rm PQ}}{N}=\frac{v_{\sigma}}{\sqrt{2}N} \ ,
\end{align}
where $N$ denotes the color anomaly factor. A nonzero color anomaly factor $N$ is essential for the axion to couple to gluons and thereby provide a consistent solution to the strong CP problem. For our model the $N$ is given as follows ($\omega$ is $\rm PQ$ charge of $\Psi$),
\begin{align}
    N&= 2 \sum_f \left( \omega^f_L- \omega^f_R \right) T (R_f) \nonumber \\
    &= 2 (q+ \frac{1}{2}- q + q+ \frac{1}{2} -q) T(1,0) \nonumber \\
    &= 1, \quad \quad T(1,0) \equiv \frac{1}{2}. 
\end{align}
This value of the color anomaly factor remains consistent with the original KSVZ axion models~\cite{Kim:1979if,Shifman:1979if}. 
We also note that the residual \(Z_3\) symmetry ensuring the Dirac nature of neutrinos is distinct from the axion domain-wall number, which is determined
by the QCD anomaly and the periodicity of the axion potential.
The QCD axion mass, obtained from the non-perturbative potential at next-to-leading order (NLO), is given as follows \cite{GrillidiCortona:2015jxo},
\begin{align} \label{eq:axionmass}
    m_a = 5.70(7) \left( \frac{10^{12}\ \text{GeV}}{f_a} \right) \mu\text{eV}.
\end{align}
This relation between $m_a$ and $f_a$ is a model independent prediction of the QCD axion, provided that the PQ symmetry is explicitly broken only by non-perturbative QCD effects. In the following subsection, we discuss how  $f_a$ is determined from DM relic density constraints, which in turn predicts the value of $m_a$.

\subsection{Axion Dark Matter} \label{sec:axiondm}
In addition to resolving the strong CP problem, the axion is an attractive DM candidate: it is light, very weakly interacting, and stable on cosmological timescales. In the early Universe, axions are typically produced nonthermally, yielding a cold DM component.
A substantial fraction of the axion DM abundance can be generated via the misalignment mechanism~\cite{Preskill:1982cy,Abbott:1982af,Dine:1982ah}, which is given by~\cite{DiLuzio:2020wdo}
\begin{equation}
\Omega_{a,\rm mis}h^2 \simeq \Omega_{\rm CDM}h^2\,
\langle \theta_0^2 \rangle \left(\frac{f_a}{2\times 10^{11}\ {\rm GeV}}\right)^{7/6},
\label{eq:misalrelic}
\end{equation}
where $\theta_0$ denotes the initial misalignment angle and $\langle \theta_0^2\rangle$ depends on whether the PQ symmetry is broken before or after inflation.

If the PQ symmetry is broken after inflation, the initial misalignment angle $\theta_0$ varies randomly between causally disconnected patches, and one averages over $\theta_0$ with the periodic
axion potential (including anharmonicities), giving $\langle\theta_0^2\rangle \sim 2.15^2$~\cite{Visinelli:2009zm,DiLuzio:2020wdo}.
Assuming that axions produced via misalignment account for the entire cold DM abundance, Eq.~\eqref{eq:misalrelic} yields $f_a \simeq 5.4\times 10^{10}\ {\rm GeV}$, leading to $m_a \simeq 106\ \mu{\rm eV}$
via Eq.~\eqref{eq:axionmass}.

Moreover, the formation and decay of cosmic strings and domain walls can substantially enhance the axion relic abundance in the post-inflationary scenario. Accounting for these contributions, the total relic density is given by~\cite{Kawasaki:2014sqa}
\begin{equation}
\Omega_{a,\rm tot}h^2 \simeq (1.6 \pm 0.4)\times 10^{-2}
\left(\frac{f_a}{10^{10}\ {\rm GeV}}\right)^{(6+n)/(4+n)},
\end{equation}
with the QCD scale fixed at $400\ {\rm MeV}$ and $n=6.68$. 
These contributions effectively shift the value of $f_a$ required to reproduce the observed DM abundance and thus imply a correspondingly shifted axion mass range. Requiring that
$\Omega_{a,\rm tot}h^2$ accounts for $100\%$ of cold DM, we obtain
\begin{equation}
f_a \simeq (4.5-7.0)\times 10^{10}\ {\rm GeV}, \qquad
m_a \simeq (81-127)\ \mu{\rm eV}.
\label{eq:favalue}
\end{equation}
For this case, we assume that the domain-wall problem is absent, as in the $N_{\rm DW}=1$ case. If 
$N_{\rm DW}>1$, so a small bias term lifting the degeneracy among the vacua, or alternatively a pre-inflationary PQ-breaking history, is required to avoid stable domain walls.

If instead the PQ symmetry is broken before, or sufficiently early
during, inflation and is not restored afterwards, inflation
homogenizes the axion field over our observable Universe. The initial
angle $\theta_0$ is then effectively a single constant rather than a
random variable, and one should replace
$\langle\theta_0^2\rangle\to\theta_0^2$ in
Eq.~\eqref{eq:misalrelic}. Topological defects produced before
inflation are inflated outside our observable patch. Consequently, the
DM-motivated value of $f_a$, and hence $m_a$, becomes $\theta_0$-dependent, and the axion parameter space is described by
a continuous band rather than a narrow interval. This realization is subject to the usual axion isocurvature constraint.
We assume that the inflationary Hubble scale and the homogeneous initial
angle $\theta_0$ satisfy this constraint.

The present model contains colored exotic scalars and vector-like colored fermions. Depending on the mass ordering, the lightest colored exotic state can be stable or long-lived at the renormalizable level.
For the best-fit points in Tables~\ref{tab:BFNH} and \ref{tab:BFIH}, the colored scalar mass eigenstates are heavier than at least one colored fermion, and the Yukawa interactions in Eq.~\eqref{eq:yuklag} open decay channels into a colored fermion accompanied by a lepton or a right handed neutrino. The lightest colored fermion, however, may remain stable in the minimal renormalizable setup.

If produced with an appreciable abundance in the early Universe, such a long-lived colored particle would survive until the QCD confinement epoch and form exotic heavy hadrons, which are subject to stringent cosmological and terrestrial constraints. An alternative possibility would be to introduce UV-suppressed operators that remove the accidental stability of the lightest colored state and induce its decay well before BBN. Since the realization of such operators depends on the ultraviolet completion, we do not invoke this possibility here.
Instead, we adopt a pre-inflationary PQ-breaking cosmology in which the PQ symmetry is not restored after inflation and
\begin{equation}
T_{\rm osc}^{a,{\rm RD}} < T_{\rm RH}
\leq T_{\max}
\ll M_{\rm C}^{\rm min},
\qquad
T_{\max}<T_{\rm PQ}^{\rm rest},
\label{eq:colored_reheating}
\end{equation}
where $T_{\rm osc}^{a,{\rm RD}}$ is the axion oscillation temperature evaluated assuming radiation domination, $T_{\rm RH}$ is the reheating temperature, $T_{\max}$ is the maximum temperature attained by the thermal plasma during reheating, $M_{\rm C}^{\rm min}$ is the mass of the lightest colored exotic state, and $T_{\rm PQ}^{\rm rest}$ denotes the model dependent thermal PQ-restoration temperature. The lower bound on $T_{\rm RH}$ ensures that reheating is completed before the onset of the standard axion oscillations, so that the relic density expressions presented above remain applicable. The upper bounds are chosen such that the thermal production of the colored exotics is exponentially suppressed to a cosmologically negligible level and the PQ symmetry is not thermally restored.
These conditions admit a consistent reheating window provided that the axion oscillation scale lies below both the maximum temperature compatible with a negligible colored exotic yield and the PQ-restoration temperature. The precise extent of this window is model dependent and would require a dedicated analysis of the reheating dynamics and the production and annihilation of the colored exotics, which is beyond the scope of this work.

We further assume that direct decays of the inflaton or the PQ radial mode into the colored exotic sector have negligible branching fractions, and that production through preheating, scatterings involving hard nonthermal inflaton decay products, or gravitational effects remains negligible. We also assume that the post-inflationary evolution of the PQ-breaking fields does not induce nonthermal restoration of the PQ symmetry or nonadiabatic production of the colored exotics. Under these assumptions, any population of colored exotics present before inflation is exponentially diluted, while no appreciable population is regenerated thermally or nonthermally after inflation. Consequently, no cosmologically relevant population of colored exotics is present at the QCD confinement epoch, and the associated constraints from exotic heavy hadrons, BBN, and terrestrial searches for anomalous heavy isotopes are avoided within the specified cosmological history.

Finally, while the standard misalignment mechanism~\cite{Preskill:1982cy,Abbott:1982af,Dine:1982ah} has long been regarded as the primary source of axion DM, recent developments have highlighted the kinetic misalignment mechanism~\cite{Co:2019jts}, where the initial axion velocity $\dot a_0$ (or kinetic energy) can play a crucial role. In this scenario, the axion field evolves from initial conditions defined by a nonzero $\dot a_0$ and a misalignment angle $\theta_0$. Once the axion field begins coherent oscillations, it behaves as cold DM, and its relic density can be estimated accordingly. If the kinetic energy density exceeds the axion potential barrier at a given temperature, the onset of oscillations is delayed until the kinetic energy becomes subdominant, potentially modifying the final relic abundance.

\subsection{Axion-Photon Coupling} \label{sec:apc}
The axion experimental program spans a variety of complementary detection strategies, with helioscopes and haloscopes serving as key probes. Together with indirect constraints from astrophysical and cosmological observations, axions are typically probed primarily through the axion-photon coupling. This coupling arises in the effective Lagrangian
\begin{equation}
\mathcal{L}=\frac{g_{a\gamma\gamma}}{4}\,a\,F_{\mu\nu}\tilde F^{\mu\nu},
\end{equation}
where $g_{a\gamma\gamma}$ denotes the axion-photon coupling, $F_{\mu\nu}$ is the electromagnetic field strength, and $\tilde F^{\mu\nu}$ is its dual. Utilizing NLO chiral Lagrangian techniques~\cite{GrillidiCortona:2015jxo}, the axion-photon coupling is given by
\begin{equation}
g_{a\gamma\gamma}=\frac{\alpha_e}{2\pi f_a}
\left[\frac{E}{N}-1.92(4)\right],
\label{gagg}
\end{equation}
where, $\alpha_e \approx 1/137$ is the fine structure constant.
Within our model framework, since the only chiral fermions with nonvanishing PQ charges are the SU(2)$_L$ singlet fields $\Psi_i$, the ratio of the electromagnetic and color anomaly factors is $E/N=0$.
Thus Eq.~\eqref{gagg} simplifies to
\begin{equation}
g_{a\gamma\gamma}=-1.92(4)\,\frac{\alpha_e}{2\pi f_a}.
\label{gaggs}
\end{equation}
Using the post-inflationary relic density motivated range of $f_a$ in Eq.~\eqref{eq:favalue}, the axion-photon coupling and axion mass lie in the ranges,
$|g_{a\gamma\gamma}| \simeq (3-5)\times 10^{-14}\ {\rm GeV}^{-1}, ~ m_a \simeq (81-127)\ \mu{\rm eV}$, respectively. This provides a definite target region for the axion searches in the post-inflationary cosmological history discussed in Sec.~\ref{sec:axiondm}.

In the pre-inflationary scenario considered for the minimal colored exotics, Eq.~\eqref{gaggs} remains valid, but $f_a$ is determined by the (unknown) single value of $\theta_0$ in our Hubble patch through Eq.~\eqref{eq:misalrelic}. Therefore, the model predicts a continuous band in the $(m_a,\,g_{a\gamma\gamma})$ plane rather than a narrow interval. In this sense, Eq.~\eqref{eq:favalue} should be viewed as a concrete post-inflationary benchmark, while the pre-inflationary case maps to a broader $\theta_0$-dependent region. 

\begin{figure} [!h]
    \centering
    \includegraphics[width=0.75\textwidth]{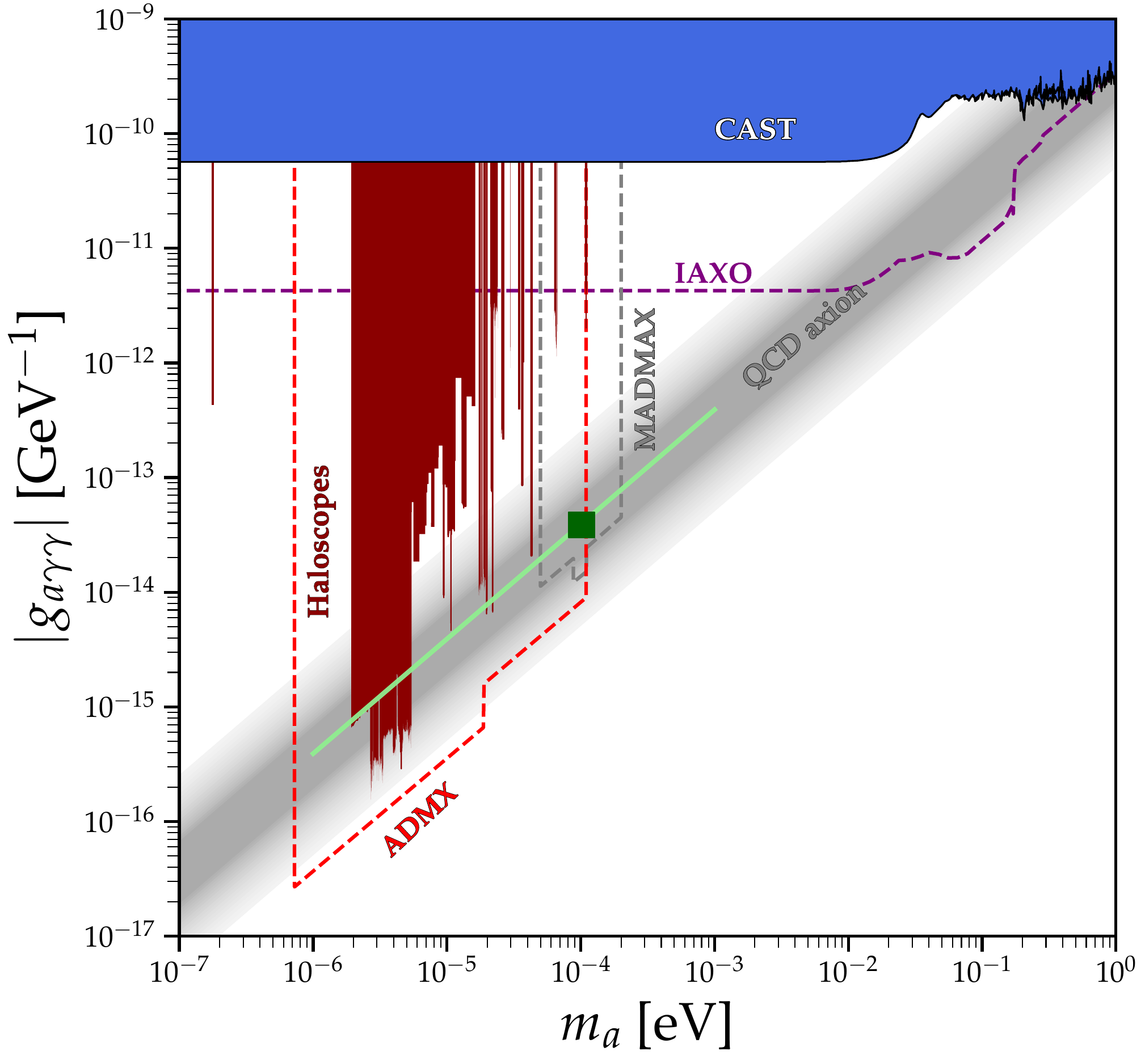}
    \caption{Axion parameter space in the plane of axion mass $m_a$ and axion-photon coupling $|g_{a\gamma\gamma}|$. The 
     post-inflationary prediction of the model is shown by the dark green patch, which lies within the QCD axion window indicated by the gray-shaded region \cite{DiLuzio:2016sbl}. The light green line corresponds to an illustrative cosmology-dependent extension of the KSVZ-like axion parameter space, with $E/N=0$.
     The current exclusion limits from haloscope experiments \cite{ADMX:2018gho,ADMX:2019uok,ADMX:2021nhd,CAPP:2020utb,Crisosto:2019fcj,HAYSTAC:2020kwv,HAYSTAC:2023cam,HAYSTAC:2024jch,Quiskamp:2024oet,QUAX:2024fut} and the CAST helioscope experiment \cite{CAST:2017uph} are represented by the dark red and blue shaded regions, respectively. The projected future sensitivities of IAXO \cite{Armengaud:2014gea,6473834}, ADMX \cite{Stern:2016bbw}, and MADMAX \cite{Beurthey:2020yuq} are shown by the dashed purple, red, and gray regions, respectively.}
    \label{fig:axion}
\end{figure}
In Fig.~\ref{fig:axion}, we show the axion parameter space in the plane of axion mass $m_a$ and axion-photon coupling $|g_{a\gamma\gamma}|$. As discussed above, the post-inflationary prediction of the model follows from Eq.~\eqref{eq:favalue}. The corresponding parameter space is shown by the dark green patch in Fig.~\ref{fig:axion}. The allowed QCD axion window is indicated by the gray-shaded band \cite{DiLuzio:2016sbl}. Since our framework corresponds to a KSVZ-type axion model, the predicted parameter space lies within this region. In addition, the light green line illustrates the broader $\theta_0$-dependent region possible in the pre-inflationary scenario, along the same KSVZ-like QCD axion line with $E/N=0$.
The dark red shaded regions denote the exclusion limits from various haloscope experiments, including ADMX \cite{ADMX:2018gho,ADMX:2019uok,ADMX:2021nhd}, CAPP \cite{CAPP:2020utb}, and others \cite{Crisosto:2019fcj,HAYSTAC:2020kwv,HAYSTAC:2023cam,HAYSTAC:2024jch,Quiskamp:2024oet,QUAX:2024fut}. In addition, the blue shaded region represents the exclusion bound from the CAST helioscope experiment \cite{CAST:2017uph}.
Upcoming helioscope and haloscope searches, including IAXO~\cite{Armengaud:2014gea,6473834}, ADMX~\cite{Stern:2016bbw}, and MADMAX~\cite{Beurthey:2020yuq}, 
will significantly extend the sensitivity to unexplored regions of axion parameter space, probing smaller values of $|g_{a \gamma \gamma}|$ over a broad range of axion masses $m_a$ and strengthening constraints on axion DM models. In particular, IAXO is projected to reach sensitivities down to $|g_{a \gamma \gamma}|$ down to $(10^{-12}-10^{-11})$ $\text{GeV}^{-1}$, as indicated by the dashed purple region. Similarly, the mass regions $1 \ \mu \text{eV} \lesssim m_{a} \lesssim 100 \ \mu \text{eV}$ and $50 \ \mu \text{eV} \lesssim m_{a} \lesssim 120 \ \mu \text{eV}$ are expected to be probed by ADMX and MADMAX, respectively. The corresponding future sensitivity regions are shown by the dashed red and gray bands.
The predicted mass range overlaps with the target region of future
high-frequency haloscope searches, in particular MADMAX-like experiments.
Because the model predicts a KSVZ-like suppressed axion-photon coupling,
helioscope searches such as IAXO are less directly sensitive to the central
benchmark region. Thus, the most direct probes of the post-inflationary benchmark are expected to come from next-generation haloscope searches.

Indirect probes of axion DM models arise from astrophysical and cosmological observations, where axion DM may decay or convert into photons in regions with intense magnetic fields or high matter densities~\cite{DiLuzio:2020wdo,OHare:2024nmr}. Such indirect probes span a wide range of axion masses, from the ultralight regime ($m_a \lesssim 10^{-14} \ \text{eV}$) to masses well above the MeV scale. Our current model framework, lies in the mass region $m_a \sim \mathcal{O}(10-100) \ \mu \text{eV}$. The Breakthrough Listen project, using radio surveys in the C-band, has searched for axion to photon conversion in the magnetospheres of neutron stars near the Galactic Center. These searches place upper limits on the axion-photon coupling $g_{a\gamma\gamma} \lesssim 10^{-11} \ \text{GeV}^{-1}$ for axion masses in the range $15 \ \mu\text{eV} \lesssim m_a \lesssim 35 \ \mu\text{eV}$~\cite{Foster:2022fxn}. Likewise, the analyses of radio data obtained from the Bullet Cluster (1E 0657-55.8) observations constrain $g_{a\gamma\gamma} \sim (10^{-12}-10^{-11}) \ \text{GeV}^{-1}$ for $10 \ \mu\text{eV} \lesssim m_a \lesssim 30 \ \mu\text{eV}$~\cite{Chan:2021gjl}. The presence of axions can significantly modify stellar evolution by enhancing cooling processes, such as the Primakoff effect. In particular, observations of horizontal branch stars in globular clusters impose a strong upper limit of $g_{a\gamma\gamma} \lesssim 6.6 \times 10^{-11} \ \text{GeV}^{-1}$~\cite{Ayala:2014pea}. This is consistent with limits obtained from the CAST helioscope, which places comparable limits on $g_{a \gamma \gamma}$ for $m_a \lesssim 0.02$ eV~\cite{CAST:2017uph}. Furthermore, updated limits from SN 1987A gamma ray observations have tightened constraints on the $g_{a\gamma\gamma}$ by including axion to photon conversion in the magnetic field of the supernova progenitor itself, yielding $g_{a\gamma\gamma} \lesssim 10^{-11} \ \text{GeV}^{-1}$ for axion mass in the range $1 \ \mu\text{eV} \lesssim m_a \lesssim 10^3 \ \mu\text{eV}$~\cite{Manzari:2024jns}. 
These results correspond to an order of magnitude improvement over earlier bounds, which relied solely on axion to photon conversion in the Galactic magnetic field. In our model, the predicted axion-photon coupling lies well below current experimental and astrophysical bounds, ensuring consistency with all existing constraints.
In addition, as pointed out in~\cite{Candon:2025sdm}, QCD axions with masses around $m_a \simeq 10^{-4}$ eV, which is favored by our model, might be probed by Fermi-LAT or a similar gamma-ray satellite, provided the next Galactic SN is of Type Ibc.

\section{Dependence of ${\text{c}}$LFV and Lepton $g-2$ on the axion scale and neutrino sectors} \label{sec:lfv}

Extensions of the SM that account for neutrino mass generation, particularly low-scale constructions~\cite{Akhmedov:1995vm,Malinsky:2005bi,Mohapatra:1986bd,Gonzalez-Garcia:1988okv}, are often accompanied by sizable charged lepton flavor violating (cLFV) processes as well. This has led to extensive experimental efforts~\cite{Jodidio:1986mz,MEG:2016leq,COMET:2018auw,Belle:2021ysv,Moritsu:2022lem,Xing:2022rob,MEGII:2023ltw,Perrevoort:2024qtc,Haxton:2024ecp,Palo:2025oyq} dedicated to probing rare cLFV processes, particularly muon and tau decays. In our model, cLFV processes are induced by the Yukawa interactions $\overline{L}_{\alpha} \Psi_{R_j} \tilde{\eta}$, and the corresponding interaction Lagrangian can be written as
\begin{align}
    \mathcal{L} &\supset \   \alpha_{\nu}Y^{(6)}_{\mathbf{1}} \overline{e}  \Psi_{R_1} \tilde{\eta} +\gamma_{\nu}\left( Y^{(6)}_{\mathbf{2},1} \overline{\mu} + Y^{(6)}_{\mathbf{2},2} \overline{\tau} \right)\Psi_{R_1} \tilde{\eta} \nonumber \\ &+ \ \beta_{\nu}Y^{(6)}_{\mathbf{1'}} \overline{e} \Psi_{R_{2}} \tilde{\eta}  + \delta_{\nu}\left( Y^{(6)}_{\mathbf{2},1} \overline{\tau} - Y^{(6)}_{\mathbf{2},2} \overline{\mu} \right)\Psi_{R_2} \tilde{\eta} \ .
\end{align}
\begin{figure}[!h]
    \centering
    \begin{tikzpicture}
\begin{feynman}
\vertex (a1) {\(\ell_{\alpha} \)};
\vertex[right=2.0cm of a1] (a2);
\vertex[below=0.05cm of a2] (a3) ;
\vertex[right=2.0cm of a2] (g22);
\vertex[below=0.2cm of g22] (g222) {\(\Psi_k\)};
\vertex[right=0.2cm of g22] (g33);
\vertex[right=2.0cm of g33] (g2) ;
\vertex[below=0.05cm of g2] (g4)  ;
\vertex[right=2.0cm of g2] (g3) {\(\ell_{\beta}  \)} ;
\vertex[above=2.1cm of g22] (e1) ;
\vertex[above=1.5cm of e1] (e2) ;
\vertex[below=0.05cm of e1] (e22) ;
\vertex[above=0.05cm of e2] (e5) {\(\gamma\)};

\diagram* {
(a1) -- [fermion, black,very thick] (a2) -- [fermion,black,very thick] (g22),
(g22) -- [fermion,black,very thick] (g2),
(g2) -- [fermion, black,very thick, edge label' =\hspace{1cm} ] (g3), 
(a2) -- [scalar,black,very thick, quarter left,edge label=\( \eta^{\pm} \)] (e1) -- [scalar,black,very thick,quarter left,edge label=\(\eta^{\pm} \)] (g2),
(e1) -- [ boson,black,very thick] (e2),
};
\end{feynman}
\end{tikzpicture}
    \caption{\centering One loop Feynman diagram for the radiative cLFV processes $\ell_{\alpha} \rightarrow \ell_{\beta} \gamma$.}
    \label{fig:lfvloop}
\end{figure}
We consider the radiative cLFV processes $\ell_{\alpha} \rightarrow \ell_{\beta}  \gamma$ ($\equiv (\mu \rightarrow e \gamma), \ (\tau \rightarrow e \gamma), \ (\tau \rightarrow \mu \gamma)$) as shown in Fig.~\ref{fig:lfvloop}, and the corresponding branching ratio ($\mathcal{BR}$) is given by
\begin{align} \label{eq:br}
    &\mathcal{BR}(\ell_{\alpha} \rightarrow \ell_{\beta} \gamma) \approx \mathcal{BR}(\ell_{\alpha} \rightarrow \ell_{\beta} \nu_{\alpha} \overline{\nu_{\beta}}) \times \frac{3 \alpha_e N_c^2}{16 \pi G^2_F}\left| \sum_{k=1}^2 \mathcal{Y}_{\beta k} \mathcal{Y}^\dagger_{k \alpha} \mathcal{F} \left(M_{k}, m_{\eta^\pm} \right) \right|^2 \ ,  \\
    & \mathcal{F} \left( x,y\right) \approx \frac{2 x^6 + 3 x^4 y^2 - 6 x^2 y^4 + y^6 -12 x^4 y^2 \ln{\left( \frac{x}{y}\right)}}{12 (x^2-y^2)^4}, \\
    & \mathcal{BR}(\mu \rightarrow e \nu_{\mu} \overline{\nu_e}) \approx 1, \quad \mathcal{BR}(\tau \rightarrow e \nu_{\tau} \overline{\nu_e}) \approx 0.1785, \quad \mathcal{BR}(\tau \rightarrow \mu \nu_{\tau} \overline{\nu_{\mu}}) \approx 0.1737  . 
\end{align}
Here, $\alpha_e \approx 1/137$ is the fine structure constant, $N_c$ is color factor, Fermi constant $G_F = 1.166 \times 10^{-5}$ $\text{GeV}^{-2}$, and the matrix of Yukawa couplings $\mathcal{Y}$ is provided in Eq.~\eqref{eq:DirYuk2}.

In the neutrino sector, we have defined the following relation,
\begin{align}
\kappa_\nu\equiv 
 \frac{N_c \sin{\theta} \cos{\theta}}{16 \pi^2} \alpha_\nu \alpha'_\nu m_0, \label{eq:1}
\end{align}
where $m_0\equiv  Y^{(4)}_{\mathbf{1}} \frac{v_{\sigma}}{\sqrt{2}} =  Y^{(4)}_{\mathbf{1}}  f_a$.
While $\kappa_\nu$ is numerically constrained by neutrino oscillation data, the parameters $\sin\theta$, $\alpha_\nu$, and $\alpha'_\nu$ remain free.
Meanwhile, the range of $m_0$ is determined by the axion sector from the corresponding value of $f_a$ obtained in Eq.~\eqref{eq:favalue},
\begin{align}
f_a=(4.5-7.0)\times 10^{10}\ {\rm GeV}. \label{eq:fa_value}
\end{align}
To facilitate our analysis, we consider the following two benchmark points (BPs):
\begin{align} \label{eq:fvalues}
 f_a&=7.0\times 10^{10}\ {\rm GeV},\quad (m_a\approx 81\ {\rm \mu eV}), 
\nonumber \\
f_a&=4.5\times 10^{10}\ {\rm GeV},\quad (m_a\approx 127\ {\rm \mu eV}).
\end{align}
We can rewrite the Eq.~(\ref{eq:1}) as follows:
\begin{align}
 \alpha'_\nu =\frac{16 \pi^2}{N_c \sin{\theta} \cos{\theta}} \frac{\kappa_\nu}{m_0} \frac{1}{\alpha_\nu}. \label{eq:2}
\end{align}
Once $\alpha_\nu$ and $\theta$ are numerically  specified, $\alpha'_\nu$ is uniquely determined.
Subsequently, these fixed values $\alpha_\nu$ and $\alpha'_\nu$ are substituted into the Yukawa couplings $\mathcal{Y}$ and $\mathcal{Y}'$, 
which are respectively given by
\begin{align} \label{eq:DirYuk2}
    \mathcal{Y} \equiv     \alpha_{\nu}   \tilde{\mathcal{Y}} \equiv
\alpha_{\nu}     \begin{pmatrix}
     Y^{(6)}_{\mathbf{1}} &  \tilde\beta_{\nu} Y^{(6)}_{\mathbf{1'}} \\
        \tilde \gamma_{\nu} Y^{(6)}_{\mathbf{2},1} &  -\tilde\delta_{\nu} Y^{(6)}_{\mathbf{2},2} \\
         \tilde \gamma_{\nu} Y^{(6)}_{\mathbf{2},2} &  \tilde\delta_{\nu} Y^{(6)}_{\mathbf{2},1}
    \end{pmatrix}, \quad
     \mathcal{Y}' \equiv     \alpha'_{\nu}   \tilde{\mathcal{Y}'} \equiv
\alpha'_{\nu}   \begin{pmatrix}
         Y^{(6)}_{\mathbf{1}} &  \tilde \beta'_{\nu} Y^{(4)}_{\mathbf{2},1} &\tilde \beta'_{\nu} Y^{(4)}_{\mathbf{2},2}  \\
       \tilde  \gamma'_{\nu} Y^{(6)}_{\mathbf{1'}} &  -\tilde \delta'_{\nu} Y^{(4)}_{\mathbf{2},2} & \tilde \delta'_{\nu} Y^{(4)}_{\mathbf{2},1}
    \end{pmatrix},
\end{align}
where we have defined the rescaled parameters $\tilde \beta^{(')}_\nu\equiv \beta^{(')}_\nu/\alpha^{(')}_\nu$, $\tilde \gamma^{(')}_\nu\equiv \gamma^{(')}_\nu/\alpha^{(')}_\nu$, and $\tilde \delta^{(')}_\nu\equiv \delta^{(')}_\nu/\alpha^{(')}_\nu$.
Since the other parameters are already constrained by the numerical analysis of neutrino oscillation data, we can determine the maximal absolute values for the elements of $\mathcal{Y}$ and $\mathcal{Y}'$.
Here, the Yukawa couplings are assumed to satisfy the perturbative bounds:
\begin{align}
{\rm Max}[|\mathcal{Y}|]\le 1,\quad 
{\rm Max}[|\mathcal{Y}'|]\le 1.
\end{align}
The branching ratio given in Eq.~\eqref{eq:br} can be rewritten as follows:
\begin{align} \label{eq:br2}
    &\mathcal{BR}(\ell_{\alpha} \rightarrow \ell_{\beta} \gamma) \approx \mathcal{BR}(\ell_{\alpha} \rightarrow \ell_{\beta} \nu_{\alpha} \overline{\nu_{\beta}}) \times \frac{3 \alpha_e N^2_c}{16 \pi G^2_F}
    \frac{|\alpha_\nu|^4}{|m_0|^4}
    \left| \sum_{k=1}^2 \tilde{\mathcal{Y}}_{\beta k} \tilde{\mathcal{Y}}^\dagger_{k \alpha} \mathcal{F} \left(\tilde M_{k}, \tilde m_{\eta^\pm} \right) \right|^2,
\end{align}
 where $\tilde M_{k} \equiv M_{k}/m_0$ and $\tilde m_{\eta^\pm}\equiv m_{\eta^\pm}/m_0$.
 Analogous to the radiative cLFV processes, the muon and electron $g-2$ receive loop-induced contributions in the present model. These contributions can be expressed as
\begin{align}
    \Delta a_{\ell} \approx - \frac{m^2_{\ell} N_c |\alpha_{\nu}|^2}{16 \pi^2 |m_0|^2} \sum^2_{k=1} \tilde{\mathcal{Y}}_{\ell k} \tilde{\mathcal{Y}}^\dagger_{k \ell} \mathcal{F} \left(\tilde{M}_{k}, \tilde{m}_{\eta^\pm} \right),
\end{align}
where $\ell= e, \mu$. 
We evaluate the cLFV processes and lepton $g-2$ varying the input parameters randomly sampled within the following ranges:
\begin{align}
|\alpha_\nu| = [10^{-5},1],\quad
\sin\theta = [-1,1].
\end{align}
The values of the remaining parameters, such as $\tilde{\mathcal Y}^{(')},\ M_\Psi,\ m_{\eta^\pm}$, are obtained from a $\chi^2$ numerical analysis.
This analysis is performed to satisfy the neutrino oscillation data within the $5\sigma$ confidence level.
In our analysis, we find that the large values of $m_0$, calculated for the two benchmark values of $f_a$ given in Eq.~\eqref{eq:fvalues}, result in highly suppressed branching ratios of radiative cLFV and contributions to the lepton $g-2$.
The maximum values obtained from our numerical results for the NH case are summarized in Table \ref{tab:maxBR-NH}.
\begin{table}[!h]
    \setlength\tabcolsep{0.2cm}
    \begin{tabular}{c||c|c|c||c|c}
\hline
       $m_a$  &$\mathcal{BR}(\mu \rightarrow e \gamma)$   & $\mathcal{BR}(\tau \rightarrow e \gamma)$ & $\mathcal{BR}(\tau \rightarrow \mu \gamma)$ & $|\Delta a_e|$  & $|\Delta a_\mu|$ \\ 
       \hline \hline
      $81~{\rm \mu eV}$     &  $3.91\times 10^{-24} $ & $2.43\times 10^{-26}$ & $4.11\times 10^{-27}$ &$2.69\times10^{-24}$& $1.59\times 10^{-19}$ \\ \hline
      $127~{\rm \mu eV}$     &  $6.04\times 10^{-25} $ & $3.80\times 10^{-27}$ &  $6.36\times 10^{-28}$ &$1.06\times10^{-24}$& $6.27\times 10^{-20}$ \\ \hline
\hline
    \end{tabular}
    \caption{\label{tab:maxBR-NH}%
      \centering Maximum values of $\mathcal{BR}(l_{\alpha}\rightarrow l_{\beta} \gamma$) and
      $|\Delta a_{e(\mu)}|$ for the NH case.}
\end{table}
Similar to the NH case, the maximum values obtained for the IH case are summarized in Table~\ref{tab:maxBR-IH}.
\begin{table}[!h]
    \setlength\tabcolsep{0.2cm}
    \begin{tabular}{c||c|c|c||c|c}
\hline
       $m_a$  &$\mathcal{BR}(\mu \rightarrow e \gamma)$   & $\mathcal{BR}(\tau \rightarrow e \gamma)$ & $\mathcal{BR}(\tau \rightarrow \mu \gamma)$ & $|\Delta a_e|$  & $|\Delta a_\mu|$ \\ 
       \hline \hline
      $81~{\rm \mu eV}$     &  $2.04\times 10^{-33} $ & $2.08\times 10^{-36}$ & $1.79\times 10^{-35}$ &$7.26\times10^{-29}$& $3.08\times 10^{-24}$ \\ \hline
      $127~{\rm \mu eV}$     &  $3.64\times 10^{-34} $ & $3.70\times 10^{-37}$ &  $2.81\times 10^{-36}$ &$3.06\times10^{-29}$& $1.30\times 10^{-24}$ \\ \hline
\hline
    \end{tabular}
    \caption{\label{tab:maxBR-IH}%
      \centering Maximum values of $\mathcal{BR}(l_{\alpha}\rightarrow l_{\beta} \gamma$) and
      $|\Delta a_{e(\mu)}|$ for the IH case.}
\end{table}
From Tables~\ref{tab:maxBR-NH} and \ref{tab:maxBR-IH}, we find that the maximum values in the IH case are smaller than those in the NH case. This behavior can be understood from the best-fit values of the model parameters obtained from neutrino oscillation data, which are summarized in Tables~\ref{tab:BFNH} and \ref{tab:BFIH} for the NH and IH scenarios, respectively. As discussed earlier, in evaluating the branching ratios and $\Delta a_{\ell}$, we vary $\alpha_{\nu}$ and $\sin{\theta}$, while fixing the remaining parameters according to the neutrino sector analysis. In order to fit the neutrino oscillation data, we find that $\alpha_{\nu}$ takes smaller values in the IH case compared to the NH case. Since the branching ratios and $\Delta a_{\ell}$ scale with powers of $\alpha_{\nu}$, their values are consequently suppressed in the IH scenario relative to the NH case. 
%

In addition to the radiative cLFV decays discussed above, the model also gives rise to the axion-emitting cLFV processes $\ell_{\alpha} \rightarrow \ell_{\beta} a$, which are also induced at the one loop-level. The corresponding Feynman diagram is shown in Fig. \ref{fig:axionlfvloop}.
\begin{figure}[!h]
    \centering
    \begin{tikzpicture}
\begin{feynman}
\vertex (a1) {\(\ell_{\alpha} \)};
\vertex[right=2.0cm of a1] (a2);
\vertex[below=0.05cm of a2] (a3) ;
\vertex[right=2.0cm of a2] (g22);
\vertex[above=0.1cm of g22] (g222) {\(\Psi_k\)};
\vertex[below=1.5cm of g22] (e2) ;
\vertex[right=0.2cm of g22] (g33);
\vertex[right=2.0cm of g33] (g2) ;
\vertex[below=0.05cm of g2] (g4)  ;
\vertex[right=2.0cm of g2] (g3) {\(\ell_{\beta}  \)} ;
\vertex[above=2.1cm of g22] (e1) ;
\vertex[below=0.05cm of e1] (e22) ;
\vertex[below=0.05cm of e2] (e5) {\(a\)};

\diagram* {
(a1) -- [fermion, black,very thick] (a2) -- [fermion,black,very thick] (g22),
(g22) -- [fermion,black,very thick] (g2),
(g2) -- [fermion, black,very thick, edge label' =\hspace{1cm} ] (g3), 
(a2) -- [scalar,black,very thick, quarter left,edge label=\( \eta^{\pm} \)] (e1) -- [scalar,black,very thick,quarter left,edge label=\(\eta^{\pm} \)] (g2),
(g22) -- [ scalar,black,very thick] (e2),
};
\end{feynman}
\end{tikzpicture}
    \caption{\centering One loop Feynman diagram for the axion-emitting cLFV processes $\ell_{\alpha} \rightarrow \ell_{\beta} a$.}
    \label{fig:axionlfvloop}
\end{figure}
Although this process is controlled by the same Yukawa interactions as the radiative cLFV decay shown in Fig.~\ref{fig:lfvloop}, it is described by a different effective operator and loop function. As a result, the flavor structure remains the same while the corresponding amplitudes differ. The corresponding branching ratio for $\ell_{\alpha}\rightarrow\ell_{\beta}a$ can be expressed as \cite{Escribano:2020wua,Escribano:2021ymx,Herrero-Brocal:2023czw,CentellesChulia:2025eck}:
\begin{align} \label{eq:braxion}
    &\mathcal{BR}(\ell_{\alpha} \rightarrow \ell_{\beta} a) \approx \mathcal{BR}(\ell_{\alpha} \rightarrow \ell_{\beta} \nu_{\alpha} \overline{\nu_{\beta}}) \times \frac{3 N_c^2}{64 \pi^2 G^2_F m^2_{\alpha} f^2_a}\left| \sum_{k=1}^2 \mathcal{Y}_{\beta k} \mathcal{Y}^\dagger_{k \alpha} M^2_{k}\mathcal{G} \left(M_{k}, m_{\eta^\pm} \right) \right|^2 \ ,  \\
    & \text{where} \quad \quad  \mathcal{G} \left( x,y\right) \approx \frac{ x^2 - y^2 -2 y^2 \ln{\left( \frac{x}{y}\right)}}{(x^2-y^2)^2}, \ \text{and} \  m_{\alpha} = m_{\mu,\tau}. 
\end{align}
A naive comparison between Eqs.~\eqref{eq:braxion} and \eqref{eq:br} suggests that the branching ratio for the axion-emitting cLFV process can be enhanced relative to that of the radiative cLFV decay. 
Similar to the radiative cLFV, we calculate the branching ratio of axion-emitting cLFV for two benchmark values of $f_a$ as given in Eq. \eqref{eq:fvalues}.
The maximum values obtained for the NH and IH cases are summarized in Table \ref{tab:maxBR-NH-axion} and Table \ref{tab:maxBR-IH-axion}.
\begin{table}[!h]
    \setlength\tabcolsep{0.2cm}
    \begin{tabular}{c||c|c|c}
\hline
       $m_a$  &$\mathcal{BR}(\mu \rightarrow e a)$   & $\mathcal{BR}(\tau \rightarrow e a)$ & $\mathcal{BR}(\tau \rightarrow \mu a)$  \\ 
       \hline \hline
      $81~{\rm \mu eV}$  &  $8.89\times 10^{-13}$ &  $1.52\times 10^{-15}$  & $3.09\times 10^{-18}$   \\ \hline
      $127~{\rm \mu eV}$    &  $2.15\times 10^{-12}$ &  $3.69\times 10^{-15}$  & $7.47\times 10^{-18}$  \\ \hline
\hline
    \end{tabular}
    \caption{\label{tab:maxBR-NH-axion}%
      \centering Maximum values of $\mathcal{BR}(\ell_{\alpha}\rightarrow \ell_{\beta} a$) for the NH case.}
\end{table}
\begin{table}[!h]
    \setlength\tabcolsep{0.2cm}
    \begin{tabular}{c||c|c|c}
\hline
       $m_a$  &$\mathcal{BR}(\mu \rightarrow e a)$   & $\mathcal{BR}(\tau \rightarrow e a)$ & $\mathcal{BR}(\tau \rightarrow \mu a)$  \\ 
       \hline \hline
       $81~{\rm \mu eV}$  &  $3.98\times 10^{-13}$ &  $1.45\times 10^{-18}$  & $2.34\times 10^{-17}$   \\ \hline
      $127~{\rm \mu eV}$    &  $9.64\times 10^{-13}$ &  $3.51\times 10^{-18}$  & $5.67\times 10^{-17}$  \\ \hline
\hline
    \end{tabular}
    \caption{\label{tab:maxBR-IH-axion}%
      \centering Maximum values of $\mathcal{BR}(\ell_{\alpha}\rightarrow \ell_{\beta} a$) for the IH case.}
\end{table}
The obtained values of the branching ratio of axion-emitting cLFV remain well within the experimental limits \cite{Jodidio:1986mz,Belle-II:2022heu,Belle:2025bpu}.
Thus, the model remains consistent with the current experimental constraints on both radiative and axion-emitting cLFV, while providing a connection between the neutrino and axion sectors.

\FloatBarrier
\section{Conclusions} \label{sec:conc}
In this work, we have presented a unified axion model framework where tiny neutrino masses, the leptonic flavor structure, the strong CP problem, and dark matter can be explained utilizing a global $U(1)_{\rm PQ}$ and modular $S_3$ symmetry. This framework is implemented in a novel class of KSVZ-type axion models, where exotic colored fermions and scalars simultaneously serve as mediators for neutrino mass generation at the one loop-level. The PQ charge assignment serves multiple purposes: it forbids neutrino masses at the tree-level, and its spontaneous breaking to a residual $Z_3$ symmetry further ensures the Dirac nature of neutrinos. An SM singlet complex scalar $\sigma$ has been introduced, whose vev is responsible for the breaking of $U(1)_{\rm PQ}$ and leads to the emergence of the axion as a pseudo-Goldstone boson. For the minimal realization of the model, we include only two generations of the colored fermion $\Psi$ mediating the loop. As a result, the neutrino mass matrix is of rank two, yielding masses for two neutrinos while leaving one neutrino massless. 
Moreover, as the PQ breaking scale determines the masses of the mediating fields, it intrinsically connects the origin of neutrino masses with the PQ scale and, consequently, to the axion mass.

Our numerical analysis shows that the model successfully accommodates both NH and IH of neutrino masses, while remaining consistent with experimental observations. We perform a $\chi^2$ analysis to fit the neutrino oscillation data, fixing $\alpha_\ell \approx0.01$ throughout the analysis. In the NH scenario, no parameter points are found within the $1\sigma$ region. The entire range of $\text{Re}[\tau]$ is allowed, while $\text{Im}[\tau]$ is localized near 1.7. In contrast, for IH, both $\text{Re}[\tau]$ and $\text{Im}[\tau]$ are confined to narrow intervals, $\text{Re}[\tau] \in [-0.156,0.147]$ and $\text{Im}[\tau] \in [1.63,1.74]$. With the lightest neutrino mass being zero, the sum of neutrino masses, $\sum D_{\nu}$, is directly intertwined with $\Delta m^2_{\rm atm}$ and $\Delta m^2_{\rm sol}$. 
Consequently, we obtain $\sum D_{\nu}\simeq 58$ meV and $\sum D_{\nu}\simeq 100$ meV for the NH and IH cases, respectively. The NH case remains consistent with the cosmological bound from the combined DESI and CMB data, whereas the IH case lies outside this range but remains consistent with the Planck limit.
Likewise, the effective electron-neutrino mass is restricted to $m_\beta\simeq9~(49)$ meV for the NH (IH) scenario. These narrow ranges are generic consequences
of a rank two neutrino spectrum with one massless state. The corresponding figures illustrate these phenomenological consequences and allow a direct comparison with current experimental and cosmological constraints, whose
implications depend on the adopted cosmological framework. The model specific flavor structure is instead reflected mainly in the allowed regions of the modulus $\tau$ and in the resulting patterns of the lepton
mixing observables. The best-fit points correspond to
$\chi_{\rm min}=2.48$ for the NH case and
$\chi_{\rm min}=1.88$ for the IH case, and the associated parameter
values are summarized in Tables~\ref{tab:BFNH} and~\ref{tab:BFIH},
respectively.
Furthermore, the axion in this model provides a solution to the strong
CP problem and can also serve as a viable DM candidate. In the
pre-inflationary, non-restored PQ history adopted for the minimal model,
the DM-motivated axion parameters depend on the homogeneous initial
misalignment angle $\theta_0$ and are subject to the usual isocurvature constraint. For comparison, the illustrative $N_{\rm DW}=1$
post-inflationary benchmark considered in this work gives
$
|g_{a\gamma\gamma}|\simeq(3-5)\times10^{-14}\ {\rm GeV}^{-1},
$
which lies within the projected sensitivity of future axion searches. This benchmark is also consistent with current astrophysical and cosmological constraints.

In addition, using the best-fit parameters of the neutrino sector and two benchmark values of the axion decay constant, we calculate the radiative and axion-emitting cLFV decay rates and the contributions to the lepton anomalous magnetic moments. The resulting values remain consistent with current experimental constraints.
In summary, the model connects neutrino masses and lepton flavor with the axion solution to the strong CP problem and axion DM through an underlying modular symmetry.

\section*{Acknowledgments}
\noindent
 SKK and RK are supported by the National Research Foundation of Korea under grant NRF-2023R1A2C100609111. 
HO is supported by Zhongyuan Talent (Talent Recruitment Series) Foreign Experts Project.

\appendix
\section{$S_3$ tensor product rules and modular Yukawa construction} \label{sec:s3rule}

\subsection{$S_3$ symmetry and its tensor product}
The set of all possible permutations of three objects is described by the symmetric group $S_3$. It has $3!=6$ elements and the generators of this group are given by~\cite{Feruglio:2017spp,Kobayashi:2018vbk,Meloni:2023aru},
\begin{align}
    \rho(S)=\begin{pmatrix}
        1 & 0 \\
        0 & -1
    \end{pmatrix}, \quad  \rho(T)=  \frac{1}{2}\begin{pmatrix}
        -1 & -\sqrt{3} \\
        -\sqrt{3} & 1
    \end{pmatrix},
\end{align}
which obey the following relations
\begin{align}
 \left(\rho(S) \right)^2= \left(\rho(T) \right)^2= \left(\rho(S) \rho(T) \right)^3= \mathbb{I}.
\end{align}
The group has 3 irreducible representations, a trivial singlet $1$, a non-trivial singlet $1'$, and a doublet $2$.
The product rules are given as follows
\begin{align} \label{eq:S3multirule}
     1' \otimes 1'=1,  \ 1' \otimes 2 = 2, \ 2 \otimes 2 = 1 \oplus 1' \oplus 2 .
\end{align}
Given two singlets $x$, $y$ and two doublets $a= (a_1, a_2)^T$, $b= (b_1, b_2)^T$, their tensor product rules following Eq. \eqref{eq:S3multirule} can be expressed as follows~\cite{Feruglio:2017spp,Kobayashi:2018vbk,Meloni:2023aru}
\begin{align} \label{eq:S3multirule1}
    &(x)_{1'} \otimes (y)_{1'} = (x y)_{1}, \quad (x)_{1'} \otimes \begin{pmatrix}
        a_1 \\
        a_2
    \end{pmatrix}_2= \begin{pmatrix}
        - x a_2 \\
       x a_1
    \end{pmatrix}_2 \nonumber \\
    & \begin{pmatrix}
        a_1 \\
        a_2
    \end{pmatrix}_2 \otimes \begin{pmatrix}
        b_1 \\
        b_2
    \end{pmatrix}_2=  (a_1 b_1 + a_2 b_2)_1 \oplus (a_1 b_2 - a_2 b_1)_{1'} \oplus \begin{pmatrix}
        - a_1 b_1 + a_2 b_2 \\
        a_1 b_2 + a_2 b_1
    \end{pmatrix}_2.
\end{align}

\subsection{Modular Yukawa based on $S_3$}
Here, we provide the discussion of the modular Yukawa construction in the context of $S_3$ symmetry. The modular group $\bar\Gamma$ attains a linear fractional transformation $\gamma$, acting on the modulus $\tau$ in the upper-half complex plane. The transformation is defined as follows
\begin{equation}\label{eq:tau-SL2Z}
\tau \longrightarrow \gamma\tau= \frac{a\tau + b}{c \tau + d}\ , \quad \text{where} \quad  a,b,c,d \in \mathbb{Z}, \quad \text{and} \quad ad-bc=1, \quad \text{Im}[\tau]>0.
\end{equation}
This is isomorphic to the transformation $PSL(2,\mathbb{Z})=SL(2,\mathbb{Z})/\{I,-I\}$. The modular transformation is generated by two fundamental operations, $S$ and $T$ given by
\begin{eqnarray}
S:\tau \longrightarrow -\frac{1}{\tau}\ , \qquad\qquad
T:\tau \longrightarrow \tau + 1\ ,
\end{eqnarray}
These operations satisfy the algebraic relations,
$S^2 =\mathbb{I}$, and $(ST)^3 =\mathbb{I}$.
We consider a series of groups, denoted as $\Gamma(N)$ for $N=1,2,3,\dots$, defined as follows
\begin{align}
\Gamma(N)= \left \{ 
\begin{pmatrix}
a & b  \\
c & d  
\end{pmatrix} \in SL(2,\mathbb{Z})~ \bigg| ~
\begin{pmatrix}
a & b  \\
c & d  
\end{pmatrix} \equiv
\begin{pmatrix}
1 & 0  \\
0 & 1  
\end{pmatrix} ~~({\rm mod}~ N) \right \}.
\end{align}
For $N=2$, we denote $\bar\Gamma(2) \equiv \Gamma(2)/\{I,-I\}$. Since the element $-I$ is not in $\Gamma(N)$ for $N>2$, one can have $\bar\Gamma(N) = \Gamma(N)$, which are infinite normal subgroups of $\bar\Gamma$, referred to as principal congruence subgroups.
The quotient groups, defined as $\Gamma_N \equiv \bar\Gamma/\bar\Gamma(N)$, come from finite modular groups. For these finite groups, the condition $T^N=\mathbb{I}$ is imposed. Specifically, the groups $\Gamma_N$ with $N=2,3,4,5$ are isomorphic to $S_3$, $A_4$, $S_4$, and $A_5$, respectively \cite{deAdelhartToorop:2011re}.
Modular forms of level $N$ are holomorphic functions $f(\tau)$, which transform under the action of $\Gamma(N)$ as
\begin{equation}
f(\gamma\tau)= (c\tau+d)^{k_I} f(\tau)~, ~~ \gamma \in \Gamma(N)~ ,
\end{equation}
where $k_I\ge0$ is referred to as the modular weight. For $k_I=0$, the modular form is a constant.  In this work, we discussed the group $S_3$, which corresponds to $N=2$.

The group $S_3$ exhibits three irreducible representations: a trivial
singlet $1$, a non-trivial singlet $1'$, and a doublet $2$. The lowest modular weight is 2 and the corresponding modular coupling is represented as $Y^{(2)}_{\mathbf{2}} \equiv \left(Y^{(2)}_{\mathbf{2},1}(\tau), \ Y^{(2)}_{\mathbf{2},2} (\tau) \right)^T \equiv \left( y_1(\tau), \ y_2 (\tau) \right)^T$. In terms of Dedekind eta-function $\eta(\tau)$, it can be expressed as~\cite{Kobayashi:2018vbk} 
\begin{align}
    &y_1(\tau)= \frac{i}{4 \pi} \left(\frac{\eta'{(\tau/2)}}{\eta{(\tau/2)}} +\frac{\eta'{((\tau+1)/2)}}{\eta{((\tau+1)/2)}} - \frac{8\eta'{(2\tau)}}{\eta{(2\tau)}} \right), \nonumber \\
    & y_2(\tau)= \frac{\sqrt{3}i}{4 \pi} \left(\frac{\eta'{(\tau/2)}}{\eta{(\tau/2)}} - \frac{\eta'{((\tau+1)/2)}}{\eta{((\tau+1)/2)}} \right),
\end{align}
where $\eta'(\tau)$ is first derivative w.r.t. $\tau$ and the expression of Dedekind eta-function $\eta (\tau)$ is given by
\begin{align}
    \eta(\tau)= q^{1/24} \prod_{n=1}^{\infty} (1-q^n), \quad q \equiv e^{2i\pi\tau}.
\end{align}
In the form of $q$-expansion, we can write modular Yukawas as follows ($y_i(\tau) \equiv y_i$)
\begin{align} 
y_1 &= \frac 18\left(1 + 24q + 24q^2 + 96 q^3 + 24q^4 + 144q^5  + 96 q^6 + 192q^7 + 24q^8 + 312q^9 + 144q^{10}+\cdots \right),\nonumber \\
y_2 &= \sqrt 3 q^{1/2} (1+ 4 q + 6 q^{2} + 8 q^{3} + 13 q^{4} + 12 q^{5} + 14 q^{6} + 24 q^{7} + 18 q^{8} + 20 q^{9} +     \cdots  ).
\label{eq:qexp}
\end{align} 
The higher modular weight Yukawas can be constructed from weight $2$ Yukawas $y_1$, $y_2$, following the tensor product rules of $S_3$ provided in Eq.~\eqref{eq:S3multirule1}. The modular Yukawas of weight $4$ and $6$ are obtained as follows
\begin{align}
    &Y_\mathbf{1}^{(4)}= y_1^2 + y_2^2, \quad  Y_\mathbf{1'}^{(4)}= 0, \quad  Y_\mathbf{2}^{(4)}= \begin{pmatrix}
        -y_1^2 + y_2^2 \\
        2 y_1 y_2
    \end{pmatrix} , \nonumber \\
    & Y_\mathbf{1}^{(6)}= -y_1^3 + 3 y_1 y_2^2, \quad  Y_\mathbf{1'}^{(6)}= -3 y_1^2 y_2 + y_2^3, \quad Y_\mathbf{2}^{(6)}=\begin{pmatrix}
        y_1^3 + y_1 y_2^2 \\
        y_1^2 y_2 + y_2^3
    \end{pmatrix}.
\end{align}
In a similar manner, the other higher modular weight Yukawa can be constructed following the tensor product rules of $S_3$ given in Eq.~\eqref{eq:S3multirule1}.

\bibliographystyle{utphys}
\bibliography{references}             
\end{document}